\documentclass[10pt,3p,final,twocolumn]{elsarticle}

\usepackage{graphicx}

\usepackage{amssymb}
\usepackage{amsthm}
\usepackage{upgreek}
\usepackage{float}
\usepackage{amsmath}
\usepackage{subfig}
\usepackage{array}
\usepackage{url}
\usepackage[para,online,flushleft]{threeparttable}

\biboptions{sort&compress}

\journal{Nuclear Instruments and Methods A}

\begin{document}

\begin{frontmatter}

\author[labelTRI]{A.B.~Garnsworthy\corref{cor1}}
\ead{garns@triumf.ca}
\cortext[cor1]{Corresponding author: 4004 Wesbrook Mall, Vancouver, BC, V6T 2A3, Canada}

\title{The GRIFFIN Data Acquisition System}

\author[labelTRI]{C.J.~Pearson}
\author[labelTRI]{D.~Bishop}
\author[labelTRI]{B.~Shaw}
\author[labelTRI]{J.K.~Smith}
\author[labelTRI]{M.~Bowry}
\author[labelGuelph]{V.~Bildstein}
\author[labelTRI]{G.~Hackman}
\author[labelGuelph]{P.E.~Garrett}
\author[labelTRI]{Y.~Linn}
\author[labelMon]{J-P.~Martin}
\author[labelTRI]{W.J.~Mills}
\author[labelGuelph]{C.E.~Svensson}

\address[labelTRI]{Physical Sciences Division, TRIUMF, 4004 Wesbrook Mall, Vancouver, BC, Canada, V6T 2A3}

\address[labelGuelph]{Department of Physics, University of Guelph, Guelph, ON, Canada, N1G 2W1}

\address[labelMon]{Department of Physics,
  Universit{\' e} de Montr{\' e}al, Montr{\' e}al, QC, Canada}

\begin{abstract}
Gamma-Ray Infrastructure For Fundamental Investigations of Nuclei, GRIFFIN, is a new experimental facility for radioactive decay studies at the TRIUMF-ISAC laboratory. This article describes the details of the custom designed GRIFFIN digital data acquisition system. The features of the system that will enable high-precision half-life and branching ratio measurements with levels of uncertainty better than 0.05\% are described. The system has demonstrated the ability to effectively collect signals from High-purity germanium crystals at counting rates up to 50~kHz while maintaining good energy resolution, detection efficiency and spectral quality.
\end{abstract}

\begin{keyword}
Digital electronics \sep digital signal processing \sep HPGe \sep TRIUMF \sep ISAC
\end{keyword}

\end{frontmatter}


\section{Introduction}
\label{sec:intro}

Large arrays of detectors for gamma-ray measurements coupled with auxiliary particle detection systems provide one of the most powerful and versatile tools for studying exotic nuclei through nuclear spectroscopy at rare-isotope beam facilities.
Gamma-Ray Infrastructure For Fundamental Investigations of Nuclei, GRIFFIN \cite{Svensson14}, is a new experimental facility for radioactive decay studies at the TRIUMF-ISAC laboratory \cite{Dilling14} located in Vancouver, Canada.

GRIFFIN will be used for decay spectroscopy research with low-energy radioactive ion beams. A high gamma-ray detection efficiency provided by an array of sixteen High-Purity Germanium (HPGe) clover detectors \cite{Rizwan16}, combined with the rare-isotope beams from ISAC, will support a broad program of research in the areas of nuclear structure, nuclear astrophysics, and fundamental interactions.

The data acquisition (DAQ) system for GRIFFIN takes advantage of custom-designed digital electronics with many of the concepts based on the DAQ System \cite{Martin08} developed for the TRIUMF-ISAC Gamma-Ray Escape-Suppressed Spectrometer (TIGRESS) array \cite{Hackman14}. The main design goals for the system were to facilitate the requirements of the two main themes of the physics program for nuclear structure and precision measurements relevant to the study of Fermi superallowed beta decays. They are to;
\begin{itemize}
\item operate at high data throughput with each HPGe crystal counting at a rate of up to 50 kHz, and
\item achieve a level of accountability and deadtime/event traceability consistent with half-life and branching ratio measurements to a precision better than 0.05 \%.
\end{itemize}
This article describes the hardware and firmware aspects of the new digital data acquisition system which have been developed to meet and exceed these design goals. Section \ref{sec:DetProps} describes the detector signal properties that will be processed by the GRIFFIN DAQ system. Section \ref{sec:System} provides an overview and description of the hardware components of the system. A detailed discussion of the signal processing algorithms applied in the firmware is provided in Section \ref{sec:SignalProcess}, followed in Section \ref{sec:Data} by a description of the information that these algorithms provide in the offline data. The user interface to the system is detailed in Section \ref{sec:UserInterface}. The performance of the system for spectroscopic measurements is examined in Section \ref{sec:Performance} and the features necessary for performing precision measurements discussed in Section \ref{sec:PrecMeas}. Finally Section \ref{sec:Future} presents a look towards future development plans and a summary is given in Section \ref{sec:summary}.

\section{Detector Properties}
\label{sec:DetProps}

\begin{table*}[htp]
\begin{center}
\caption{\label{tab:Det_Sigs}Typical properties of the detector signals which are processed by the GRIFFIN DAQ system.}
\begin{tabular}{lccccc}
\hline
Detector & Detector & Risetime & Decay constant & Amplitude & Polarity \\
Name & Type & (ns) & ($\mu$s) & (V) & \\
\hline
HPGe+Preamp & Semiconductor & 300 & 50 & 0.1-1 & Negative\\
Si(Li)+Preamp & Semiconductor & 200 & 8 & 0.1-1 & Negative\\
BGO+Preamp & Scintillator & 400 & 0.8 & 0.1-1 & Positive\\
Plastic & Scintillator & 6 & 0.018 & 0.1-0.3 & Negative\\
Plastic+Preamp & Scintillator & 100 & 20 & 0.1-1 & Negative\\
BC422Q Plastic & Scintillator & 1 & 0.006 & 0.1-0.3 & Negative\\
LaBr$_3$ & Scintillator & 8 & 0.06 & 0.1-1 & Negative\\
DESCANT & Scintillator & 10 & 0.06 & 0.1-10 & Negative\\
\hline
\end{tabular}
\end{center}
\end{table*}%

The signals from various radiation detector types will be processed by the GRIFFIN DAQ system and as such a variety of digitizer types are required. The signal properties of the primary detectors are shown in Table \ref{tab:Det_Sigs}. Two classes of radiation detectors are considered; semiconductors and scintillators which have relatively slow and fast signals respectively.

In the initial implementation of the GRIFFIN DAQ system digitizers with sampling rates of 100~MHz and 1~GHz have been pursued, with 14-bit and 12-bit resolution respectively. These are combined with other modules for data readout, clock synchronization and slow-equipment controls. These modules are described in detail in the following section and are capable of handling the requirements for semiconductor and scintillator type detectors.

\section{System Overview and Hardware Description}
\label{sec:System}

The MIDAS (Maximum Integrated Data Acquisition System) data acquisition system \cite{MIDAS} developed at PSI (Paul Scherrer Institute, Switzerland) and TRIUMF has been adopted to interface with the GRIFFIN electronics modules. This provides a framework for the interface with the electronics modules as well as a high-throughput system for writing experimental data to disk storage.

\begin{figure*}
\centering
\includegraphics[width=1.0\linewidth]{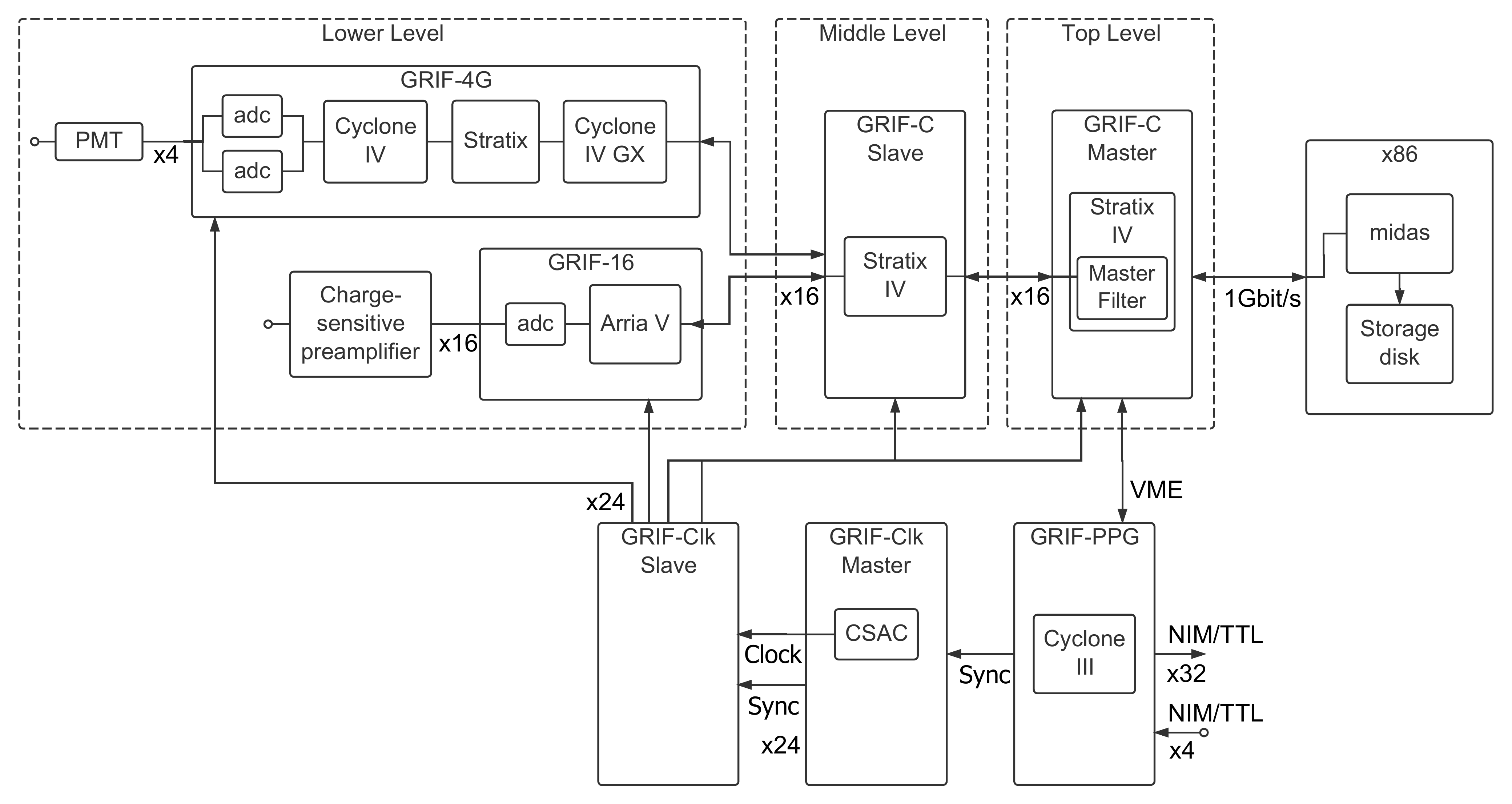}
\caption{Schematic layout of the data and clock paths within the GRIFFIN DAQ system.}
\label{fig:DAQ_Schematic}
\end{figure*}

The hardware of the GRIFFIN DAQ system is arranged into three levels. A schematic representation of the system is presented in Figure \ref{fig:DAQ_Schematic}. The lower level is the digitizer modules, the GRIF-16 and GRIF-4G modules, which process the detector signals. The middle level contains GRIF-C Slave collector modules that collect the data from all lower-level digitizers and communicate it to the top level. Each GRIF-C Slave module concentrates the 16 input data links into a single output data link. At the top level is the GRIF-C Master collector module which filters the data from all modules at the lower level to either reject it or accept it for read-out to the MIDAS frontend. Also associated with the top level is the GRIF-Clk Master clock module providing the reference clock distributed to all modules through GRIF-Clk Slave modules, as well as the GRIF-PPG programmable pulse generator module used to output analogue control signals for external equipment synchronization.

All modules in the system are single-width, 6U VME (Versa Module Europa) modules and are operated in VME64x crates \cite{VME}. The upper, P1 VME connector is used to provide power to the module and for VME communication. Only the GRIF-C Master and the GRIF-PPG modules communicate over the VME communication bus. The lower, P2 connector is reserved for analogue signal feedthrough from the rear-transition cage, as discussed in the description of the GRIF-16 module (See Section \ref{sec:GRIF-16}).

Data communication between modules is performed over high-quality 4-to-1 cables using the mini-SAS connector (SFF-8088). These cables provide one bi-directional data path between each lower-level module and the next level up.

All modules use a single clock distributed at 50~MHz from the GRIF-Clk Master module as the reference clock for all Field-Programmable Gate Array (FPGA) devices and data links. This clock signal, in addition to a synchronization signal, is provided to all modules in the system using cables with e-SATA connectors. The sync signal is used to indicate to all modules in the system the start of data collection and is used to zero all timestamps and counters.

Detailed descriptions of each module in the system are given in the remainder of this section.

\subsection{The GRIF-16 Digitizer for processing signals with a preamplifier}
\label{sec:GRIF-16}

\begin{figure}
\centering
\includegraphics[width=1.0\linewidth]{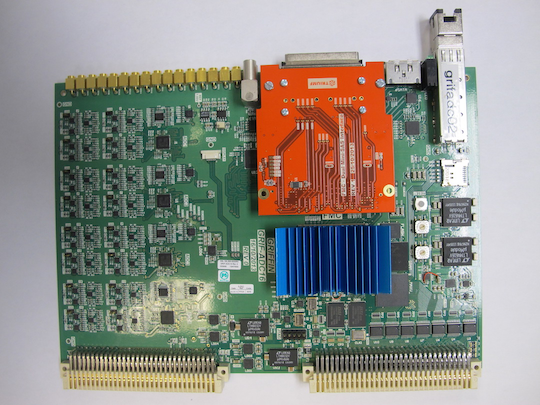}
\caption{Photograph of the GRIF-16 module.}
\label{fig:GRIF-16-photo}
\end{figure}

The GRIF-16 module (shown in Figure \ref{fig:GRIF-16-photo}) is a 16-channel, 14-bit, 100~Msamples/s analogue-to-digital converter (ADC) VME-6U module. The input signals are processed by four AD9253 quad ADCs and the signal processing of all channels is done in a single Arria-V FPGA. There is also a VITA 57 FMC (FPGA Mezzanine Card) slot \cite{FMC} to accommodate either additional input amplifiers and ADCs or additional I/O.
The GRIF-16 is used to read the signals from the GRIFFIN HPGe detectors, plastic scintillator detectors (if used with a preamplifier), and Si(Li) detectors.

The digitizer has two alternative input paths which are selectable by a physical switch on the printed-circuit board (PCB); 16 input MCX (miniature coaxial) connectors on the front panel, or inputs through a rear-transition card connected through the VME64x P2 connector. The rear-transition card can easily be exchanged to accommodate different signal amplification/attenuation, as well as a range of input connectors as additional vertical space is available. In the GRIFFIN application the rear-transition card is used with 16 SMA (SubMiniature version A) connectors. This has the additional advantage of removing a large number of cables from the front of the VME crate which makes maintenance of the digitizers much easier.

Two dual-output, 14 bit, 250~Msps digital-to-analogue converters (DAC) provide signals to LEMO connectors located on the front panel. These same LEMO connectors can alternatively be used as NIM inputs. The front panel has a mini-SAS connector for Gigabit data link to the GRIF-C module. There is also a Small Form-factor Pluggable (SFP) connector for an ethernet connection to a general network. An eSATA connection is used to input the clock and sync signals from the GRIF-Clk. The GRIF-16 includes two types of non-volatile memory, a Micro SD Flash card slot and a DDR3 128M x 32, 667~MHz DRAM.

The SFP network connection allows for monitoring, diagnostics and control independent of the MIDAS DAQ system. This is achieved through the GRIF-16 running multiple servers simultaneously within a Nios-II embedded processor on the Arria-V FPGA. In the Nios-II the GRIF-16 module runs the $\mu$COS/II operating system. Within this operating system, multiple tasks are created to handle MSCB (MIDAS Slow Control Bus), HTTP, and MIDAS communication and control. The MSCB tasks use a customized MSCB protocol library to control the devices and IO on the board, including the ADCs. There is also a `supervisor' MSCB task that allows control via the MSCB UDP MIDAS Slow Control client. The HTTP task uses the open-source Mongoose Web Server \cite{Mongoose}, and delivers static web content, JSON, and binary blobs. The MIDAS task uses the Frozen JSON library \cite{FrozenJSON} to parse and send HTTP GET requests to/from a MIDAS HTTP server, providing a link between the GRIF-16 module and MIDAS. The final task is a networking stack that implements UDP, TCP/IP and ICMP. This allows the GRIF-16 module to respond to pings, use DHCP, and provide the above web services. Further details on the main user interface page is given in Section \ref{sec:UserInterface}. 

A MAX V CPLD controller is used to configure the devices on the module. Firmware is typically loaded from non-volatile flash memory on power-up of the card. It can also be loaded to the devices or to the flash memory through a JTAG connector on the side of the module. In addition, new firmware files can be saved to the flash memory through the ethernet connection using a SFTP server running on the Nios.

The SFP connector can alternatively be used to connect the GRIF-16 module directly to a MIDAS DAQ system to operate the module in a stand-alone mode. Such a mode of operation is suitable for a small experimental setup with only a small number of channels and avoids the requirement for the GRIF-C modules. When operated in stand-alone mode the GRIF-16 can use an on-board oscillator as the reference clock which also avoids the requirement for a GRIF-Clk module.

\subsection{The GRIF-4G Digitizer for processing signals without a preamplifier}
\label{sec:GRIF-4G}

The TIG-4G module was originally designed at the Universit\'{e} de Montr\'{e}al in 2008 to process the signals from the DEuterated SCintillator Array for Neutron Tagging (DESCANT) \cite{Garrett14} detectors in the TIGRESS DAQ system \cite{Martin08}. The TIG-4G is a 4-channel, 12-bit, 1~Gsamples/s ADC VME-6U module. The 1~Gsamples/s is accomplished by an interleaving of two KAD5512P-50 ADCs.

\begin{figure}
\centering
\includegraphics[width=1.0\linewidth]{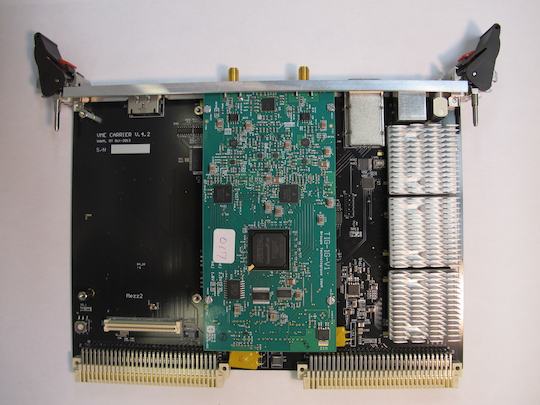}
\caption{Photograph of the GRIF-4G module. The lower mezzanine card is removed.}
\label{fig:GRIF-4G-photo}
\end{figure}

Two mezzanines accommodate two channels each with an SMA input connector for each channel. These two mezzanines share a common motherboard which accommodates the (original) Altera Stratix FPGA. Each channel uses two interleaved ADC chips each with two A/D cores sampling at 250~MHz to achieve an overall sampling rate of 1~GHz. The signal processing algorithms for the two channels of the mezzanine runs on an Altera Cyclone IV E FPGA located on the mezzanine itself.

The TIG-4G module was modified in 2014 for use in the GRIFFIN DAQ system by the addition of the eSATA clock connector (described in Section \ref{sec:GRIF-Clk}) and a mini-SAS connector with a Cyclone IV GX FPGA to enable communication with the GRIF-C module via Gigabit links. The GRIF-4G (shown in Figure \ref{fig:GRIF-4G-photo}) is used to read out the DESCANT and plastic scintillator signals directly from the PMT without the use of a preamplifier.

Firmware is loaded from non-volatile memory on power-up of the card and can be loaded by a JTAG connector on the carrier card.

\subsection{The GRIF-C Data Collector Module}
\label{sec:GRIF-C}

\begin{figure}
\centering
\includegraphics[width=1.0\linewidth]{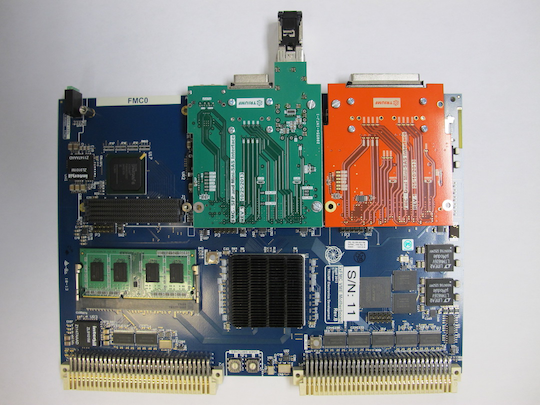}
\caption{Photograph of the GRIF-C collector module. The GRIF-C has three FMC slots, the first (centre) is populated with a Communication FMC, the second and third house a Data FMC. The third slot is empty in the photograph so the Xilinx Spartan 6 FPGA for that slot is visible. The main Altera Strarix IV GX FPGA can be seen at the centre of the carrier card.}
\label{fig:GRIF-C-photo}
\end{figure}

The GRIFFIN Collector (GRIF-C) module  (shown in Figure \ref{fig:GRIF-C-photo}) is a VME-6U module with an Altera Stratix IV GX FPGA as the main processor and a dedicated Xilinx Spartan 6 FPGA for each of the 3 FMC slots. A Micro SD Flash card slot and DDR3 128M x 32, 667~MHz DRAM memory are provided on the carrier card. In the normal GRIFFIN system the upper and lower FMC slots house a Data FMC card which has a dual-mini-SAS connector cage. These FMC cards receive data from the digitizer modules. The central FMC slot houses a Communication FMC which includes an eSATA connector for the input of the reference clock to the GRIF-C module, a mini-SAS connector for communication of data to another GRIF-C module, and a SFP connector for communication of data to the MIDAS DAQ system over a network connection. The SFP connection also allows monitoring and control communication with the GRIF-C module through several servers which run on a Nios-II embedded processor.

The collector modules are operated in two levels as shown in Figure \ref{fig:DAQ_Schematic}. The middle level collectors are connected directly to up to 16 digitizer modules each to concentrate the data collected from them and pass it to the GRIF-C Master collector module. The top level collector module is responsible for coordination of the experiment by the sequencer (see Section \ref{sec:Sequencer}), and runs the master filter algorithm (see Section \ref{sec:Filter}) to determine which data are written to disk and which are rejected.

The hardware and firmware of the GRIF-C Master and GRIF-C Slave modules are identical. A GRIF-C card will act as a Master if it is configured as such with parameter settings. A GRIF-C Master is also capable of communicating directly with the frontend digitizer cards, rather than indirectly through a GRIF-C Slave module, in order to operate in a two-level system instead of three-levels. These are useful features, especially for constructing a smaller setup where only a few digitizer cards are required.

Firmware is typically loaded from non-volatile flash memory on power-up of the card. It can also be loaded to the devices or to the flash memory through a JTAG connector on the side of the module. Alternatively new firmware files can be saved to the flash memory through a SFTP server running on the Nios-II through the ethernet connection.

Although not yet pursued in GRIFFIN, the FMC slots on the GRIF-C module are designed to also accommodate input amplifiers and ADCs in a similar manner to the GRIF-16 (described in Section \ref{sec:GRIF-16}).

\subsection{The GRIF-Clk Clock Module}
\label{sec:GRIF-Clk}

\begin{figure}
\centering
\includegraphics[width=1.0\linewidth]{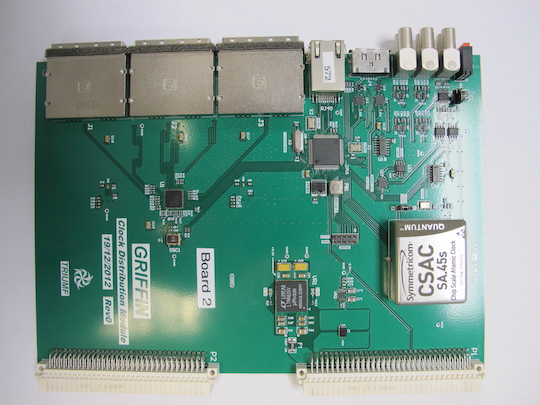}
\caption{Photograph of the GRIF-Clk Clock module. The Symmetricom Chip-Sized Atomic Clock can be seen at the lower right of the photograph.}
\label{fig:GRIF-Clk-photo}
\end{figure}

A network of GRIF-Clk clock modules (shown in Figure \ref{fig:GRIF-Clk-photo}) generate and distribute the master clock and synchronization signals for the entire GRIFFIN DAQ system. A single GRIF-Clk Master module houses a Symmetricom Model SA.45s Chip-Sized Atomic Clock (CSAC) which produces a precision 10~MHz reference signal. A 50~MHz master clock signal derived from this reference is then fanned-out through GRIF-Clk Slave modules to all collectors and digitizers in the system. Each GRIF-Clk module has 24 outputs on six 4-to-1 mini-SAS to eSATA cables. There can be 24 digitizers per GRIF-Clk Slave and 24 GRIF-Clk Slaves per GRIF-Clk Master.

In addition to the reference clock, the GRIF-Clk module distributes a sync signal which is used to indicate the start of data collection to all modules in the system and is used to zero timestamps and counters. The sync signal is generated by the sequencer code running in the GRIF-C Master which has ultimate experiment control. The signal is then output from the GRIF-PPG module (controlled by the GRIF-C Master via VME, see Section \ref{sec:GRIF-PPG}) and input to the GRIF-Clk Master. The GRIF-Clk Master distributes both the clock and sync signals to all GRIF-Clk Slave modules which then pass it to the collector and digitizer modules. When the GRIF-Clk module is operated in slave mode the input signals are simply fanned-out to the module output connectors, there are no additional clock cleaners or manipulation of the signal. As all cable lengths and hardware are identical there is very little time difference (clock skew) between the signals presented to the collector and digitizer modules. The clock skew was measured to be $<$700~ps and the skew in the sync signal is $<$400~ps. The timing jitter of this distributed clock at the input to the digitizer module was found to be $<$16~ps.

\subsection{The GRIF-PPG Programable-Pulse Generator Module}
\label{sec:GRIF-PPG}

\begin{figure}
\centering
\includegraphics[width=1.0\linewidth]{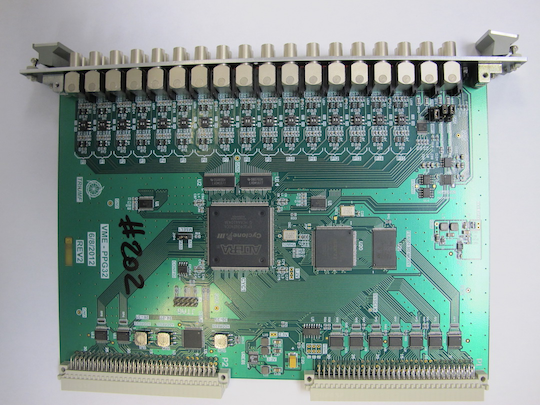}
\caption{Photograph of the GRIF-PPG Programmable Pulse Generator module.}
\label{fig:GRIF-PPG-photo}
\end{figure}

The GRIFFIN Programmable Pulse Generator (GRIF-PPG) module (shown in Figure \ref{fig:GRIF-PPG-photo}) generates NIM and TTL logic signals to be used as control signals for equipment such as the Moving-Tape collector. The front panel accommodates 32 LEMO outputs on 16 dual-LEMO connectors, each with an individual switch to select NIM or TTL. There are an additional 4 LEMO inputs, again on dual-LEMO connectors at the top of the card which are not currently utilized in the GRIFFIN DAQ setup.

The cycling of the experiment (beam-on, beam-off, tape move and operation of the electrostatic beam kicker etc) is controlled by the sequencer firmware (See section \ref{sec:Sequencer}) in the GRIF-C Master module. The GRIF-PPG module includes a Cyclone III FPGA which is more than capable of running the sequencer code directly as has been done in other experimental setups. However in the GRIFFIN system the GRIF-PPG module is used effectively as an output register. When the desired output pattern changes in the GRIF-C Master sequencer this is communicated to the GRIF-PPG module by a VME command and immediately presented on the LEMO outputs. The latency between the change in the sequencer and the LEMO output is on the order of a few microseconds which is negligible with respect to the slow equipment it is used to control in the GRIFFIN setup.

\section{Signal Processing}
\label{sec:SignalProcess}

The analogue input signal for each channel of the digitizer undergoes amplification and filtering before being converted to a digital sample by an Analogue-to-Digital Converter (ADC). A series of real-time signal processing algorithms in firmware constantly process the digitized data. The output of these algorithms are recorded when a physics event is identified in that channel by the hit detection algorithm. The details of these signal processing algorithms as implemented in the GRIF-16 digitizer are described in the following Sections, \ref{sec:SignalProcessHit} to \ref{sec:Scalers}. The equivalent algorithms in the GRIF-4G are similar in their working principles. Sections \ref{sec:Sequencer} and \ref{sec:Filter} discuss the operation of additional firmware algorithms running in the GRIF-C Master module.

In all cases the parameters of the signal processing algorithms can be easily changed through the user interface as described in section \ref{sec:UserInterface}. Typical values for the parameters are shown in Table \ref{tab:Parameters}.

\begin{table*}[htp]
\begin{center}
\caption{\label{tab:Parameters}Typical values of the parameters controlling the signal processing algorithms in the GRIF-16 module suitable for various detector signal types.}
\begin{tabular}{lllcc}
\hline
Signal Processing & Parameter & & HPGe & Plastic Scintillator\\
Algorithm & &  &  & + Preamp\\
\hline \\ [-2.0ex]
Hit detection & $T^H_{dif}$ & Differentiation length (ns) & 320 & 320 \\[1ex]
 & $T^H_{int}$ & Integration length (ns) & 80 & 80 \\[1ex]
 & $T^H_{dec}$ & Decay constant ($\mu$s) & 52.5 & 14.0 \\[1ex]
 & $T^H_{DT}$ & Imposed fixed deadtime ($\mu$s) & 1.2 & 0.3 \\[1ex]
\hline \\ [-2.0ex]
Pulse-height & $T^E_{dif}$ & Differentiation length ($\mu$s) & 8.0 & 6.0 \\[1ex]
Evaluation  & $T^E_{int}$ & Integration length ($\mu$s) & 7.0 & 4.0 \\[1ex]
 & $T^E_{del}$ & Delay before integration (ns) & 700 & 1000 \\[1ex]
 & $T^E_{dec}$ & Decay constant ($\mu$s) & 52.5 & 14.0 \\[1ex]
 & $T^E_{BLR}$ & Baseline restore speed (adc units/ns) & 0.8 & 0.8 \\[1ex]
\hline \\ [-2.0ex]
CFD timing & $T^T_{dif}$ & Differentiation length (ns) & 320 & 320 \\[1ex]
 & $T^T_{int}$ & Integration length (ns) & 10 & 10 \\[1ex]
 & $T^T_{del}$ & CFD delay length (ns) & 30 & 240 \\[1ex]
 & $F^T$ & CFD scale factor & 0.125 & 0.125 \\
\hline
\end{tabular}
\end{center}
\end{table*}%

\subsection{Hit Detection}
\label{sec:SignalProcessHit}

The hit detection algorithm identifies a physics event in the detector element connected to the channel and initiates data read-out of the other signal processing algorithms. In many cases it is desirable to be able to identify very low-energy events in the face of various background noise fluctuations of a baseline. Therefore some form of filtering or averaging must be applied in order to avoid false noise triggers. The performance of a number of filter algorithms was investigated both in simulation and on real detector signals. A simple averaging algorithm was found to be the most effective and is implemented as a simple version of the pulse-height evaluation algorithm described in Section \ref{sec:SignalProcessEnergy} but running very different parameter values as shown in Table \ref{tab:Parameters}. The method is the same as the typical moving-window deconvolution technique \cite{Georgiev93, Jordanov94, Stein96}.

The hit detection is implemented as a set of three Finite Impulse Response (FIR) digital filters,

\begin{subequations}
\begin{align}
A^H_n &= v_n - v_{(n-T^H_{dif})} \label{Eqn:HitA} \\
B^H_n &= \frac{1}{T^H_{dec}} \sum_{i=1}^{T^H_{dif}}  A^H_{(n-i)} \label{Eqn:HitB} \\
C^H_n &= \frac{1}{T^H_{int}} \sum_{i=0}^{T^H_{int}-1}  A^H_{(n-i)} + B^H_{(n-i)} \label{Eqn:HitC}
\end{align}
\end{subequations}

\noindent where $v_n$ is the digitized ADC sample at time $n$ and the constant parameters $T^H_{dif}$, $T^H_{dec}$ and $T^H_{int}$ are the differentiation window width, preamplifier decay constant and the integration time. A hit is detected when the value of $C^H_n$ is above a threshold value set by the user. An example of the result of this algorithm is shown in panel b of Figure \ref{fig:Energy_LowRate}. The sample at which this hit is detected is used as the timestamp value for the event and is equivalent to leading-edge timing. The timestamp value is the time-zero used as a reference for the other signal processing algorithms.

The typical differentiation and integration periods for the hit detection of HPGe signals as shown in Table \ref{tab:Parameters} are 320 and 80~ns respectively which allows a threshold of $\sim 5$keV (See Section \ref{sec:Thresholds}). The minimum time difference between hits for them to be distinguished is one sample longer than the differentiation time, $T^H_{dif}$ plus the risetime of the signal.

Unfortunately the risetime of HPGe pulses is not constant. A consequence of this is that the time the hit detection pulse remains above the threshold value is both energy and risetime dependent. In the case of the GRIFFIN HPGe signals from a $^{152}$Eu source it is found that the overall average risetime, measured between 10 and 90\% of the full amplitude, for all energies is $160\pm50$~ns. In the case of 121~keV pulses the risetime is $210\pm^{80}_{20}$~ns and for 1408~keV pulses it is $150\pm^{60}_{20}$~ns. At each energy the distribution of risetimes does not have a gaussian shape and displays a significant tailing on the longer-risetime side. In order to accurately know the time during which triggers can be accepted a fixed deadtime period, $T^H_{DT}$, that is much longer than the $T^H_{dif}$ plus the longest risetime time can be imposed on the hit detection algorithm. An accurate knowledge of the deadtime is essential for the considerations discussed in Sections \ref{sec:Deadtime} and \ref{sec:HighRateEfficiency} on deadtime-corrections and high-rate performance respectively.

\subsection{Pulse-Height Evaluation}
\label{sec:SignalProcessEnergy}

The pulse-height of the detector signal is usually equivalent to the energy deposited in the detector element. The pulse-height evaluation algorithm transforms the digitized signal into a trapezoidal pulse from which the pulse-height of the signal can be extracted. The method here is a slightly modified version of the usual moving-window deconvolution technique \cite{Georgiev93, Jordanov94, Stein96} and includes a correction for any slow baseline-shift which is essential for good energy resolution at high counting rates. The method is implemented as a set of four FIR digital filters,

\begin{subequations}
\begin{align}
A^E_n &= v_n - v_{(n-T^E_{dif})} \label{Eqn:EnergyA} \\
B^E_n &= \frac{1}{T^E_{dec}} \sum_{i=1}^{T^E_{dif}}  A^E_{(n-i)} \label{Eqn:EnergyB} \\
C^E_n &= C^E_{(n-1)} - T^E_{BLR} ~\text{sgn}(A^E_{(n-1)} + B^E_{(n-1)}) \label{Eqn:EnergyC}\\
D^E_n &= \frac{1}{T^E_{int}} \sum_{i=0}^{T^E_{int}-1}  A^E_{(n-i)} + B^E_{(n-i)} + C^E_{(n-i)} \label{Eqn:EnergyD}
\end{align}
\end{subequations}

where $v_n$ is the digitized ADC sample at time $n$ and the constant parameters $T^E_{dif}$, $T^E_{dec}$, $T^E_{int}$ and $T^E_{BLR}$ are the differentiation window width, preamplifier decay constant, integration time and baseline-restore speed respectively. Typical values for these parameters are shown in Table \ref{tab:Parameters}.

\begin{figure}
\centering
\includegraphics[width=0.9\linewidth]{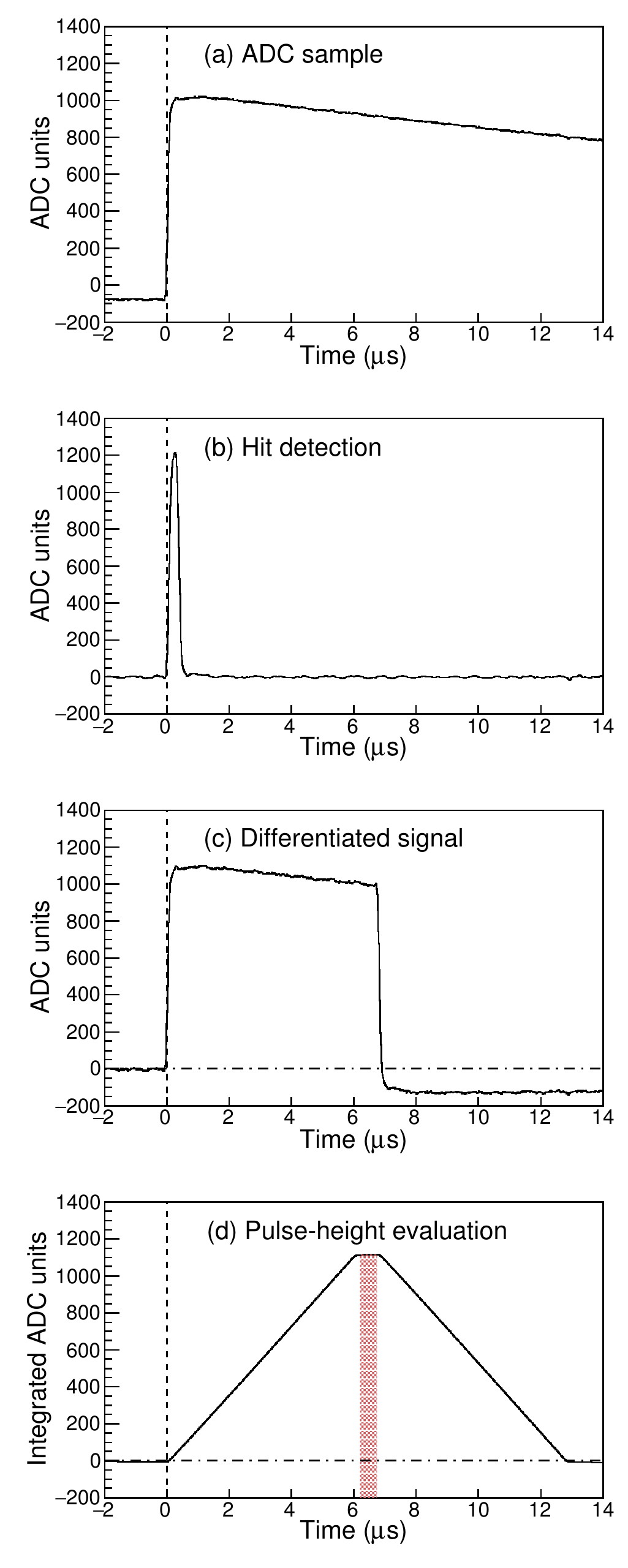}
\caption{A digitized HPGe preamplifier waveform (a), with the result of the hit detection algorithm (b) and the pulse-height evaluation algorithm (d). Panel c shows the differentiated signal (Equation \ref{Eqn:EnergyB}) at the intermediate stage of the pulse-height evaluation algorithm. The shaded red region indicates from which times following the hit detection, the final pulse-height evaluation result can be taken from. See text for details.}
\label{fig:Energy_LowRate}
\end{figure}

The filters are applied sequentially to the digitized preamplifier signal (Figure \ref{fig:Energy_LowRate}a), beginning with the difference filter given by Equation \ref{Eqn:EnergyA}. This removes the DC offset, effectively shifting the baseline of the signal to zero. After applying this filter the signal still has a negative pole associated with the decay constant, $T^E_{dec}$, of the preamplifier signal (Figure \ref{fig:Energy_LowRate}c). 
To compensate for the decaying baseline, Eq. \ref{Eqn:EnergyB} is applied to the resulting waveform to calculate the deficit associated with the decay. The deficit is added to the result of Equation \ref{Eqn:EnergyA} giving the trapezoidal signal, shown in Figure \ref{fig:Energy_LowRate}(d). The rising part of the signal has a shape which reflects the rise of the preamplifier signal and the falling part is its mirror. 
In the ideal case the preamplifier will have a single exponential component (the decay constant $T^E_{dec}$) and where this exactly known the baseline of the corrected signal would lie flat along the horizontal axis. Other noise contributions such as 60~Hz noise pick-up would still be present though as fluctuations around the baseline. Further compensation to the baseline shift is achieved using Equation \ref{Eqn:EnergyC}. This is similar to methods described in Ref. \cite{VanDevender14} but implemented here in firmware as a real-time algorithm.

Equation \ref{Eqn:EnergyD} is a low-pass filter which integrates over a number of samples, given by the parameter $T^E_{int}$, to remove the remaining high-frequency noise from the top of the signal. It is necessary to provide this integral value in the data on disk as it can deviate from the nominal value during the processing of pile-up events (See Section \ref{sec:Pile-up}). The pulse height, usually proportional to energy, can then be determined by dividing the integral returned from Equation \ref{Eqn:EnergyD} by the integration length, $T^E_{int}$. (Often a fraction of the integration length is used in order to maintain good channel dispersion in the energy spectrum.) The integration length is the equivalent of the shaping time employed by a traditional analogue spectroscopy amplifier \cite{Georgiev93, Jordanov94, Stein96}. The integration length $T^E_{int}$ must be less than the differentiation time $T^E_{dif}$. The pulse-height value is read out at the time when the sample number $i$ is equal to the time-zero as given by the hit detection plus $T^E_{del}+T^E_{int}$. The delay $T^E_{del}$ is necessary to ensure the full collection of charge has happened before the integration begins.

\subsection{Timing}
\label{sec:SignalProcessTime}

In addition to the timestamp value provided by the hit detection algorithm in Section \ref{sec:SignalProcessHit}, a timing algorithm from a digital implementation of a Constant-Fraction Discriminator (CFD) is provided. This is essential to compensate for the amplitude-dependent time-walk which is inherent to detectors such as HPGe. The time-walk effect is most significant for low-energy signals and can be as large as 300~ns for these large-volume HPGe crystals when a leading-edge type algorithm is applied. The constant fraction is evaluated as,

\begin{subequations}
\begin{align}
C^TF_n &= \frac{C^T_n}{F^T} - C^T_{(n-T^T_{del})} \label{Eqn:CFDa}
\end{align}
\text{where,}
\begin{align}
C^T_n &= v_n - v_{(n-T^T_{dif})}, \label{Eqn:CFDb}
\end{align}
\end{subequations}

\noindent $T^T_{del}$ is the CFD delay time, $F^T$ is the CFD scale factor and $T^T_{dif}$ is the differentiation length which must be greater than the risetime of the pulse. The zero-crossing point is when $C^TF_n =0$ and is taken as the result of the timing algorithm.

\subsection{Summing and Pile-up}
\label{sec:Pile-up}

\begin{figure*}
\centering
\includegraphics[width=\linewidth]{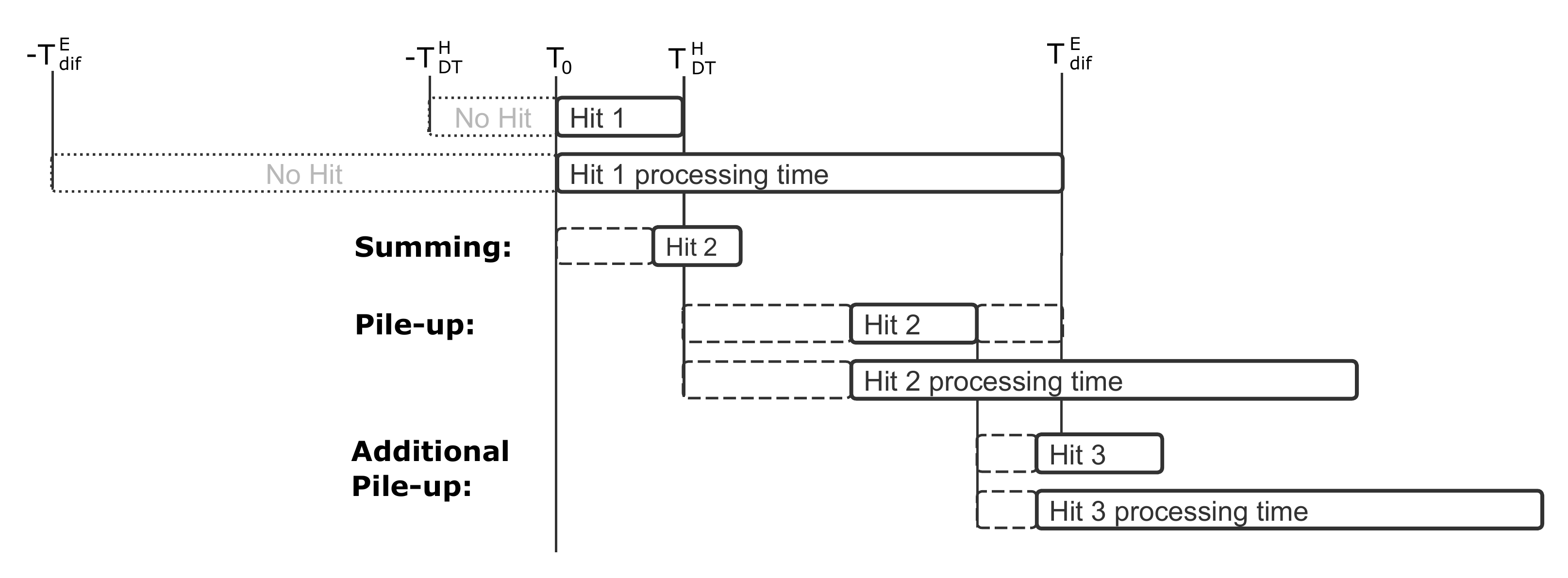}
\caption{Schematic diagram of hit detection scenarios that are processed as summing or as pile-up.}
\label{fig:Hit_Pileup}
\end{figure*}

Signal pile-up occurs when more than one energy deposition from different physics events is present in a detector element during the processing time of the initial interaction. Summing occurs when these two energy depositions occur so close in time that they cannot be distinguished. There are two distinct types of summing;
\begin{enumerate}
\item True-coincidence summing occurs when two coincident gamma rays from the same physics event are detected in the same detector element. This is dependent on the details of the nuclear decay and the geometry of the detector system, but is independent of the event rate. It will not be discussed further here.
\item Random-coincidence summing occurs when gamma rays from separate physics events are detected in the same detector elements with an unresolvable time difference. The observed energy is equal to the sum of the two energy depositions and therefore affects the apparent efficiency by removing counts from the individual photopeaks and adding counts to the sum photopeak.
\end{enumerate}
The term summing will refer to random coincidence summing in the remainder of this document. It is an important rate-dependent effect that must be taken into account in the GRIFFIN data acquisition system at the same time as considering pile-up.

Figure \ref{fig:Hit_Pileup} shows the different relative timing scenarios of hit detection events. The probability of summing is dependent on the imposed fixed deadtime, $T^H_{DT}$, of the hit detection algorithm. The probability of pile-up is dependent on the length of the differentiation window, $T^E_{dif}$, of the pulse-height algorithm which is normally 1~$\mu$s longer than the integration period, $T^E_{int}$, and on the imposed fixed deadtime, $T^H_{DT}$, of the hit detection algorithm.

\begin{figure}
\centering
\includegraphics[width=0.9\linewidth]{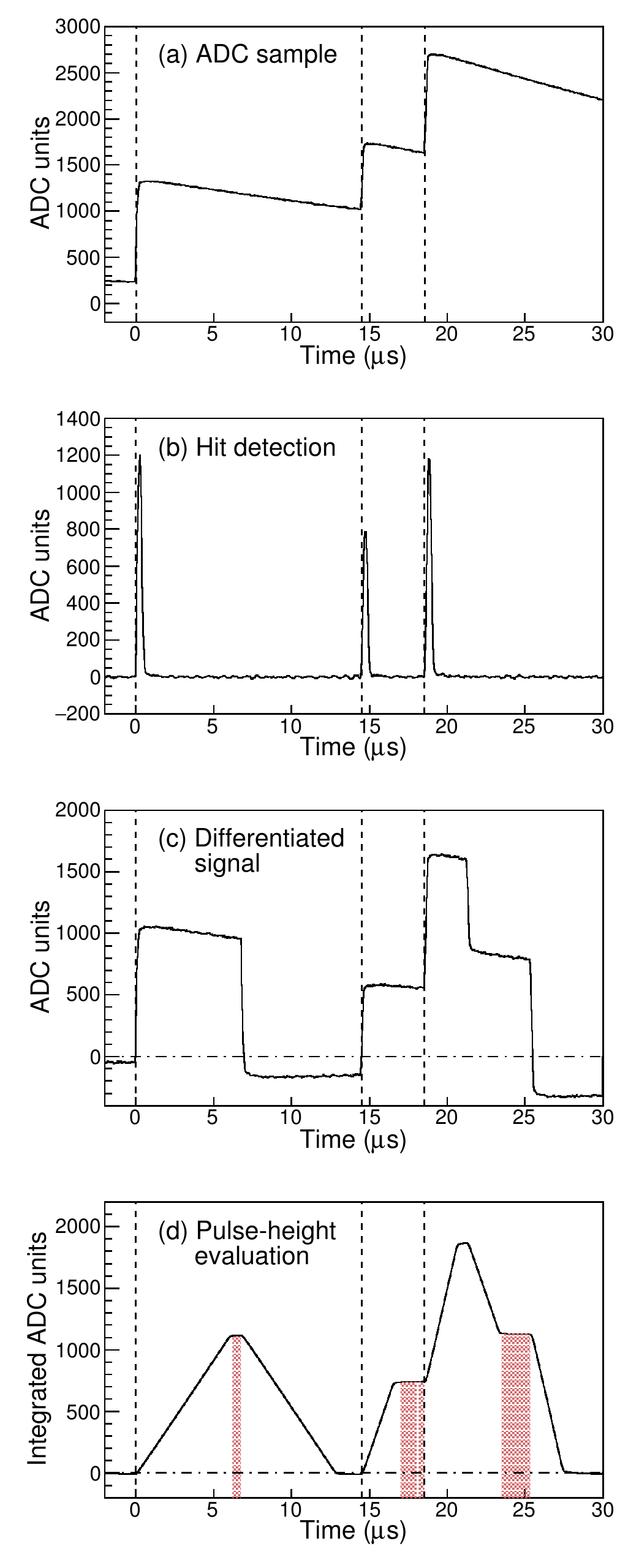}
\caption{Same as Figure \ref{fig:Energy_LowRate} but showing three interactions. The first one is a single event, the second two are considered as a pile-up event with two interactions occurring within the 6.8~$\mu$s differentiation period. The integration length used in this pile-up example is $T^E_{int}$=2~$\mu$s. The shaded red region indicates from which times following each hit detection the final pulse-height evaluation result can be taken.}
\label{fig:Energy_3pile-ups}
\end{figure}

It is a primary goal of the GRIFFIN DAQ system to operate the HPGe detectors at a counting rate up to 50~kHz per crystal. As the probability of pile-up increases with counting rate it is essential that as many events as possible are recovered so that the high efficiency of the detectors is not lost. The way in which instances of pile-up is handled in the GRIF-16 firmware is described here.

If the time between hit detections is greater than the differentiation period of the pulse-height algorithm, $T^E_{dif}$, then there is no pile-up. The original hit detection and subsequent hit detection are treated in the normal way as separate events.

If there are two or more hit detections within the differentiation period, $T^E_{dif}$, following the original hit detection (Figure \ref{fig:Energy_3pile-ups}) then these multiple hits will be handled as pile-up events. The minimum time difference between hits for them to be distinguished is one sample longer than the hit detection imposed fixed deadtime. Each hit is processed separately and generates a new event packet that will be written to disk. These events are labelled as pile-up events with a flag in the data which allows them to be treated separately in the offline analysis, if desirable.

The longest integration period possible for each hit will be used on the non-piled-up region of the signal. This will be shorter than the nominal integration period specified in the parameter for the pulse-height algorithm. The integration period used in each case, which is necessary for determining the pulse-height value for each interaction is included in the event (regardless of whether the event is pile-up or not). An improvement in the energy resolution of both hits can be made if in addition to the individual integration regions of both hits, the integral of the summed region is also recorded and a minimization performed to extract the two individual pulse-heights.

Each additional hit extends the differentiation period of the original hit. There is a limitation to the maximum number of hit detections within the original differentiation period for which reasonable energy resolution can be achieved. All events are processed but events in which an integration period of less than 1~$\mu$s was applied will have noticeably poorer energy resolution (See Section \ref{sec:Energy_Res}). This limitation is typically reached when more than 3 interactions occur within the differentiation period of the original hit. 

\begin{figure}
\centering
\includegraphics[width=\linewidth]{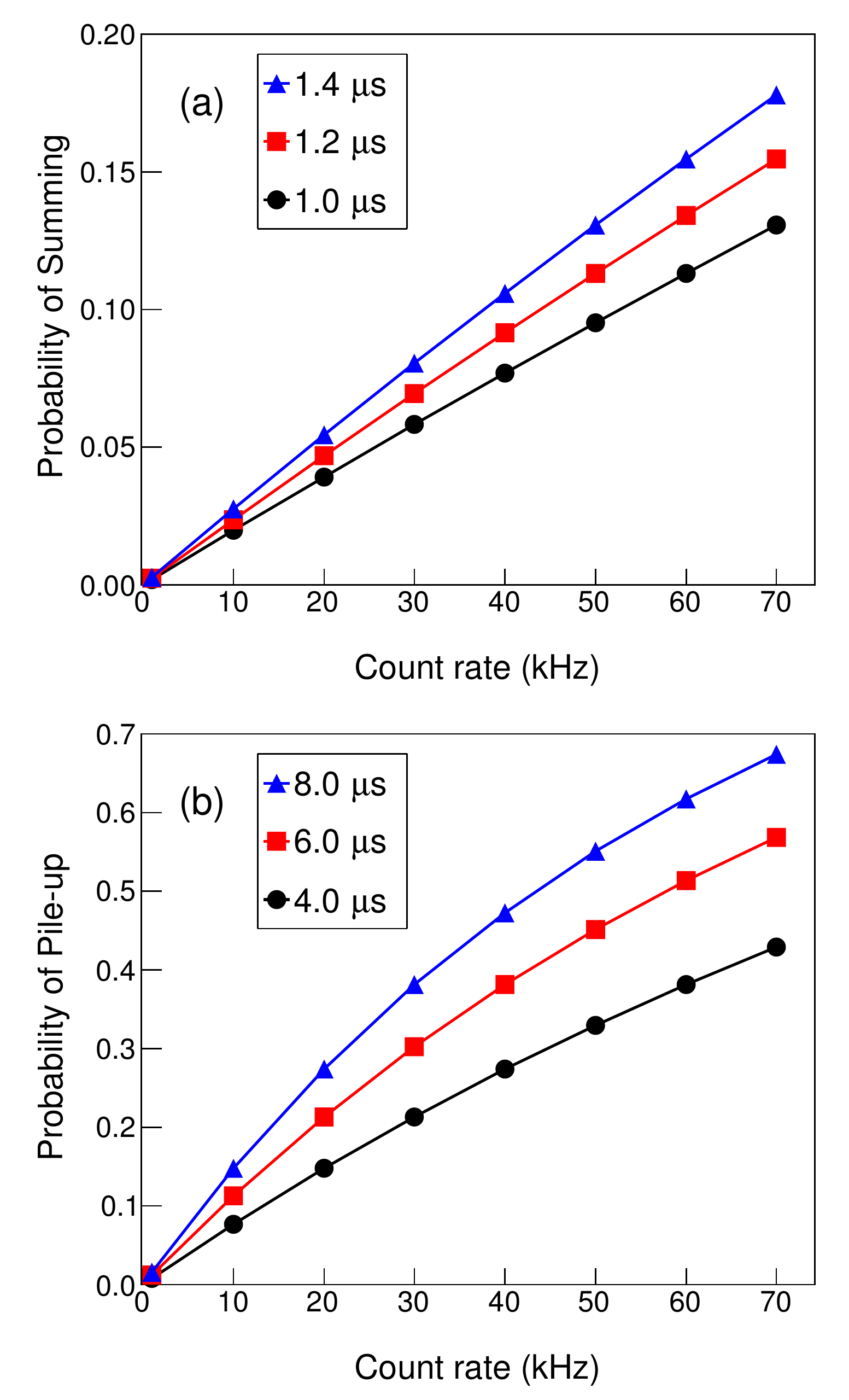}
\caption{Calculated probabilities for the occurrence of (a) summing and (b) pile-up events as a function of counting rate for different values of the imposed fixed deadtime of the hit detection and differentiation period of the pulse-height algorithms respectively.}
\label{fig:PileupFraction}
\end{figure}

The methodology and equations for calculating the probability of the various summing and pile-up scenarios in the GRIFFIN DAQ system are described in detail in \ref{sec:Append_summing_pile-up}. Based on these probabilities, it can be seen in Figure \ref{fig:PileupFraction} that the probability of summing and pile-up increases with the counting rate of the detector element. At a rate of 50~kHz with $T^H_{DT}=1.2\mu s$ and $T^E_{dif}=8\mu s$ that rejecting all pile-up will result in only 46.4\% of events being collected. The GRIFFIN DAQ system approach to pile-up recovery will recover 80.7\% of events with 10.7\% lost to $>3$ pile-ups and 8.6\% to summing. Even higher retention of relative efficiency can be achieved with short differentiation periods but at a reduction in energy resolution.

\subsection{Deadtime}
\label{sec:Deadtime}

Deadtime is the total period of time during which hit detections cannot be processed even if they are present. In the GRIFFIN DAQ system deadtime is applicable on a single channel basis and there is no overall global deadtime. System wide effects such as data loss are possible but should be treated separately from deadtime as the individual channels are still accepting and processing hits. The total deadtime period for a data collection period is equal to the imposed fixed deadtime of the hit detection algorithm multiplied by the total number of hit detections during that data-collection period, plus any additional deadtime from hits not being accepted. Hits can be rejected for instance if the memory buffer on the GRIF-16 module is full. The deadtime period since the previous accepted hit is recorded in the 14-bit deadtime word of each event. It has a maximum value of 163.83~$\mu s$.

\subsection{Event Traceability}
\label{sec:Traceability}

Events are generated for each channel in the lower-level digitizers (GRIF-16 or GRIF-4G). It is possible that events will be lost by memory buffers becoming full or link communication errors before they are written to the disk file. This event loss can be identified and accounted for by examining several counters provided in the GRIFFIN data.

The channel hit detection counter is a counter that is incremented for each individual channel each time a hit detection is identified by the hit detection algorithm. A subsequent event on disk from that channel will have a channel hit detection counter value that is one greater than the previous event unless an event was lost due to some form of data loss. If the channel hit detection counter indicates that an event was lost then it can be accounted for as effective deadtime of that channel because this 
generates a period of time, equal to the differentiation period $T^H_{dif}$, plus the signal risetime during which hit detections effectively could not be processed.
A second counter called the accepted channel hit counter is also generated in the lower-level digitizers for each individual channel. The accepted channel hit counter is only incremented when an event leaves the final memory buffer on the digitizer.
Using both the channel hit detection counter and accepted channel hit counter together can determine if any event loss has occurred in the lower-level module or in another module of the system.

An additional counter, the network packet counter, is added at the final memory buffer of the GRIF-C Master just before the event leaves the GRIFFIN electronics. The network packet counter should increment by one each time data is communicated between the GRIFFIN electronics and MIDAS. If one is missing it can be attributed to data loss over the network or in MIDAS.

Using both the accepted channel hit counter and network packet counter it is possible to identify if event loss happened in the GRIFFIN electronics or in the network transfer and MIDAS. Any event loss can be accounted for in the offline analysis of GRIFFIN data.

\subsection{Scalers}
\label{sec:Scalers}

There are two types of scalers generated by the GRIF-16 digitizers; monitoring scalers and fixed-deadtime scalers.

Scalers for monitoring the counting rates of each channel are generated in the GRIF-16 and communicated to the MIDAS system for operational monitoring purposes only and are not written to the disk file. These rates scalers reflect all identified hit detections from the hit detection algorithm for each channel.

Every channel has a set of four fixed-deadtime scalers that are read out at a regular period of time which is a user selectable parameter. Each hit detection from the hit detection algorithm is passed to the fixed-deadtime scalers. Each hit detection increments the scaler by one if the scaler is live. A different fixed deadtime is imposed on each of the scalers and subsequent hit detections are not counted if they are received within this fixed period following a hit detection. The fixed deadtimes are 0, 1, 10 and 100 microseconds. The 0$\mu$s scaler will have an actual deadtime equal to the fixed deadtime period of the hit detection algorithm.

\subsection{The Sequencer}
\label{sec:Sequencer}

GRIFFIN experiments are usually performed in a cycling mode. When the radioactive beam is delivered it is implanted into the tape of a Moving Tape collector at the central focus of the GRIFFIN detectors. When the tape moves, the sample is transported behind a wall of lead shielding and away from sight of the detectors. A typical cycle involves a tape movement, background data collection, data collection while the radioactive beam is being delivered, and finally a period of data collection with the radioactive beam blocked. Blocking of the beam is achieved by an electrostatic kicker. The relative duration of each stage within the cycle can vary between tens of milliseconds and tens of minutes in different experiments, depending on the half-lives and relative beam intensities of the isotopes involved, as well as the goal of the measurement. Control of this cycling mode is performed by the sequencer firmware running in the GRIF-C Master module.


\begin{table*}[htp]
\begin{center}
\caption{\label{tab:cycle}Parameters to operate typical measurement cycles of GRIFFIN, controlled by the sequencer firmware and providing analogue logic signals from the GRIF-PPG module.}
\begin{tabular}{lccccccc}
\hline
Operation & Bit-pattern & & \multicolumn{5}{c}{Pattern duration (s)}\\
 & & Isotope: & $^{32}$Na & $^{62}$Ga &$^{26}$Na & $^{46}$K & $^{152}$Eu\\
 & & Half life: & 13.2(4)~ms & 116.121(21)~ms & 1.07128(25)~s & 105(10)~s & Offline source\\
\hline
Move Tape & 0xC008 & & 1 & 1.5 & 1 & 1.5 & -\\
Background & 0xC002 & & - & 2 & 0.5 & 10 & -\\
Beam on& 0xC001 & & 0.2 & 10 & 3 & 150 & -\\
Beam off& 0xC004 & & 1 & 2 & 15 & 300 & $\infty$\\
\hline
\end{tabular}
\end{center}
\end{table*}%

A series of parameters are loaded to the sequencer at the beginning of data collection. The parameters represent the bit pattern of logic signals to be output on the front-panel LEMO connectors of the GRIF-PPG during each part of the cycle, and the duration for each pattern to be output. The parameters for a typical measurement cycle are shown in Table \ref{tab:cycle}. The sequencer communicates the desired bit pattern to the GRIF-PPG module through a VME communication. After the specified duration for that pattern, the next pattern is communicated to the GRIF-PPG module. This continues until the end of the data collection.

The duration is specified in microseconds so the 32-bit parameter can accommodate a maximum duration of 71.583~mins. Longer times can be constructed by repeating the same pattern in sequential parameters if desired.

The GRIF-C Master produces an event that is written to disk each time the sequencer pattern changes. This event includes the timestamp at which the change occurs, the previous pattern in the sequencer, the new pattern in the sequencer and the pattern read back from the GRIF-PPG module which should match the new pattern. These special PPG events in the disk file can be used to reconstruct the cycles during offline analysis of the data.

Although the bit patterns are 16-bits, there are only actually two logic signals required to operate a typical GRIFFIN cycle; one for the electrostatic kicker that controls beam delivery, and one to initiate a movement of the Moving Tape collector. The background collection and the period of data collection with the radioactive beam blocked are in fact an identical hardware situation. However, using a different bit pattern to indicate these different parts of the cycle is very useful in the offline data analysis. In the future, more complex hardware setups could be controlled with the other logic signals, for example lasers on/off for polarization measurements or a vacuum valve opened/closed for gas-handling as part of the measurement cycle. An example of such an experimental setup is that which would be required for an electric dipole moment search in radon isotopes at TRIUMF-ISAC \cite{Tardiff14}.

\subsection{The Master Filter}
\label{sec:Filter}

\begin{figure*}
\centering
\includegraphics[width=1.0\linewidth]{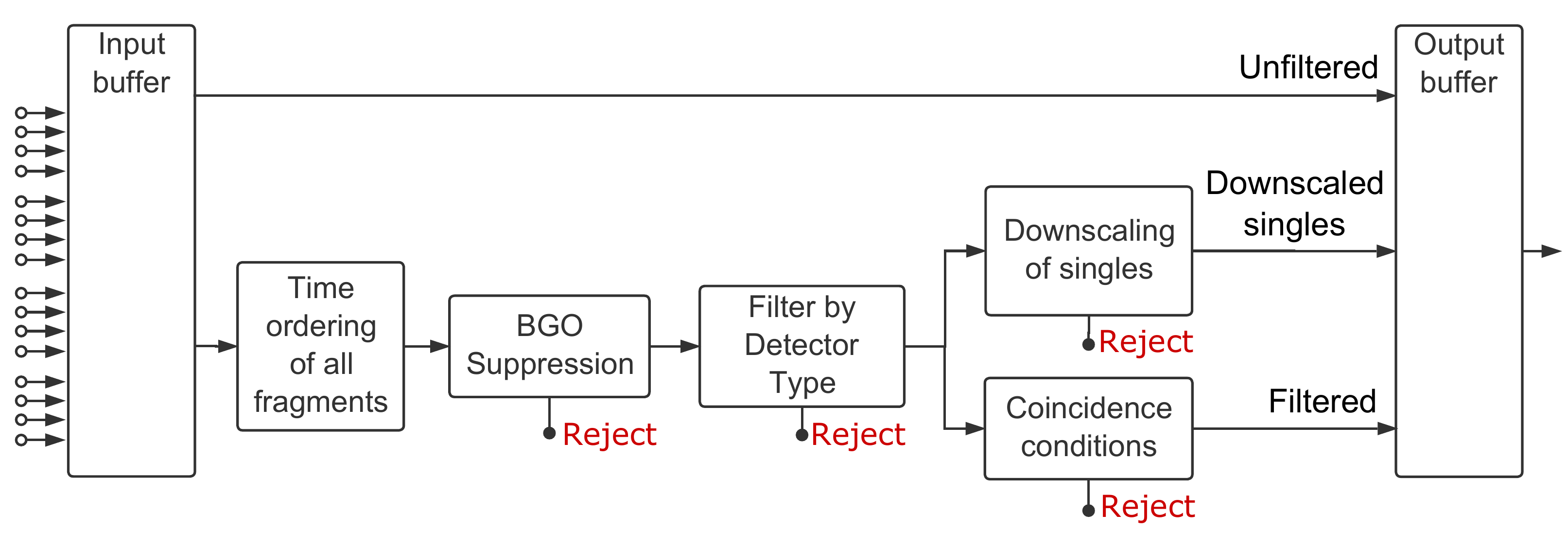}
\caption{Schematic layout of the master filter algorithm in the GRIF-C Master.}
\label{fig:Master_Filter_Schematic}
\end{figure*}

The master filter consists of a series of logic modules built in the firmware of the GRIF-C Master and executed in real time. All events collected in the digitizers are presented to the master filter where they are either rejected or accepted and written to disk. Figure \ref{fig:Master_Filter_Schematic} gives an overview of the flow of events through the master filter. The function of each component is described here.\\

\noindent {\it Input Buffer:}\\
The input buffer receives events from all GRIF-C Slave modules and must be capable of holding all events collected during the processing time of the master filter algorithm.
An {\it unfiltered} data path can be used to bypass completely the other components of the master filter. This was especially useful in the development and testing of the system.
The {\it filtered} data path proceeds through all other components of the master filter. The majority of the event data is held at the input buffer while the detector type, address and CFD-corrected timestamp are processed by the algorithms in order to determine if the event should be accepted or rejected. If an event is rejected then it is deleted from the input buffer. If an event is accepted then it proceeds to the output buffer for communication to disk storage.\\

\noindent {\it Time-ordering:}\\
The algorithms in the components that follow depend on the events being in sequential time order. There can be some latency of events originating from digitizers connected to different GRIF-C Slaves. Therefore the first stage of the master filter is to time-order the incoming events based on the CFD-corrected timestamp provided by the hit detection signal processing algorithm.\\

\noindent {\it BGO Suppression:}\\
Events which are identified with the detector type of HPGe are scrutinized for the presence of events from the associated BGO suppression shields. The mapping of BGO suppression shield channels to HPGe channels is controlled by parameters which are loaded to the GRIF-C Master at the start of each data collection period. If an associated BGO event is present within a user-defined coincidence window then the HPGe and BGO events will be rejected and will not be processed by the following components of the master filter.

This part of the master filter can optionally be disabled to ensure all HPGe and BGO events are written to disk where Compton-suppression can be performed offline.\\

\noindent {\it Detector Type Selection:}\\
A parameter is used to determine which detector types can proceed to be considered by the downscaling and coincidence condition algorithms. There are provisions for up to 14 different detector types. Any events with detector types that are not enabled are rejected at this point.\\

\noindent {\it Downscaled singles:}\\
The ability to record Downscaled singles data is essential in high counting rate situations in order to extract absolute branching ratios without overwhelming the data rate with singles data. Downscaled singles events are not presented to the Coincidence Filtering algorithm but are passed directly to the Output Buffer for communication to disk storage. As an example for Downscaled Beta events with a Downscale factor of 100, the algorithm would accept one in every 100 beta events presented to it. The one accepted event is passed to the Output Buffer and the other 99 are rejected. Using the 16-bit parameter size the Downscale factor can have a value between 2 and 65,535.\\

\noindent {\it Coincidence Filter:}\\
A series of parameters define the required multiplicity of various detector types which must be present within a user-defined coincidence window in order to pass the Coincidence Filter. Up to 14 different Coincidence Filter conditions can be set at one time. Each Coincidence Filter condition can involve one or more detector types. As an example, HPGe doubles or a $\beta$-$\gamma$-neutron coincidence use one and three detector types respectively. Singles events are also included as Coincidence Filter conditions with multiplicity 1 for that detector type. In addition, there are special detector types of ``Clover'' and ``Suppressed Clover'' that can be used in Coincidence Filter conditions.

The event CFD-corrected timestamps are used to inspect the relative time difference between events. Once a Coincidence Filter condition is passed all events in the coincidence window at that time are flagged as passing the Coincidence Filter and are subsequently passed to the Output Buffer. For this reason it is essential to control which detector types are presented to the Coincidence Filter with the detector type Selection algorithm to avoid a large data rate from a detector type that is not desired. However this behavior does ensure that if a gamma-beta event occurs then any additional coincident neutron events will pass the master filter if DESCANT is enabled. Each event is labeled with which master filter condition or conditions have been passed, and a counter representing how many events have passed the master filter since the beginning of the data collection period.
\\

\noindent {\it Output Buffer:}\\
All data that is passed to the Output Buffer is communicated over the network to the backend computer in order to be written to storage disk.

\section{Structure of the Data}
\label{sec:Data}

The data from the GRIFFIN DAQ system is processed by the MIDAS data acquisition system \cite{MIDAS} which provides a framework for writing the experimental data to disk storage. The format of the data follows the standard MIDAS approach of a series of 32-bit words wrapped by standard MIDAS header and trailer words. Table \ref{tab:DataStructure} lists all the information contained in each 352-bit basic GRIF-16 event consisting of eleven 32-bit words. The fixed-value bits used as masks to identify the purpose of data words is not included in the size value but is in the 352-bit total. If waveform samples are included in the readout then this event size is larger as there are additional words. Also if more than one filter condition has been satisfied the event size is increased with an additional word for each filter condition passed. Multiple GRIF-16 events (typically $\sim$1000) are contained within each MIDAS event on disk.

The GRIF-4G module can produce some additional information if processing the DESCANT detector type. These are not discussed in detail here but include the results of several pulse-shape analysis algorithms used to distinguish neutron and gamma events in the deuterated scintillator \cite{Bildstein15}.

\begin{table*}[htp]
\begin{center}
\caption{\label{tab:DataStructure}List of the information provided in a data event written to disk. BOR stands for Beginning Of Run, representing the start of the data collection period.}
\begin{threeparttable}
\begin{tabular}{lcl}
\hline
Data & Size\tnote{1} & Description \\
& (bits) & \\
\hline
Module type & 3 & Module type (GRIF-16 etc) that generated this event\\
Address & 16 & Unique identifier for the channel that produced this data\\
Detector type & 4 & Identifies the detector type (HPGe, BGO etc)\\
Network packet counter\tnote{2} & 28 & Total network packets generated since last module initialization\\
Filter condition counter\tnote{3} & 31 & Total number of events passed by filter since BOR\\
Filter patterns & 14 & Reports which filter patterns this event passed\\
Timestamp & 42 & Elapsed time since the BOR\\
Channel hit detection counter & 28 & Total number of hits in this channel since BOR\\
Channel accepted hit counter & 14 & Total hits in this channel while it was live since BOR\\
Pile-up type & 5 & Reports the number of hits during this event\\
Integration length & 14 & Integration length applied in the pulse-height algorithm\\
Pulse-height & 28 & Result of the pulse-height algorithm\\
CFD & 22 & Result of the timing algorithm\\
Deadtime & 14 & The duration of the deadtime preceding this event\\
Waveform & N*14 & Raw ADC data (present if waveform readout is enabled)\\
\hline
\multicolumn{3}{l}{Total size of standard event without waveforms = 32-bit $\times$ 11 words = 352 bits}\\
\hline
\end{tabular}
\begin{tablenotes}
\item[1] This size value only includes the data bits. Fixed-value bits of the 32-bit words used as masks to identify the purpose of data words is not included.\\
\item[2] Only has a value for the first GRIFFIN event in a network packet.\\
\item[3] May appear multiple times in one event.\\
\end{tablenotes}
\end{threeparttable}
\end{center}
\end{table*}%

\section{User Interface}
\label{sec:UserInterface}

\begin{figure}
\centering
\includegraphics[width=1.0\linewidth]{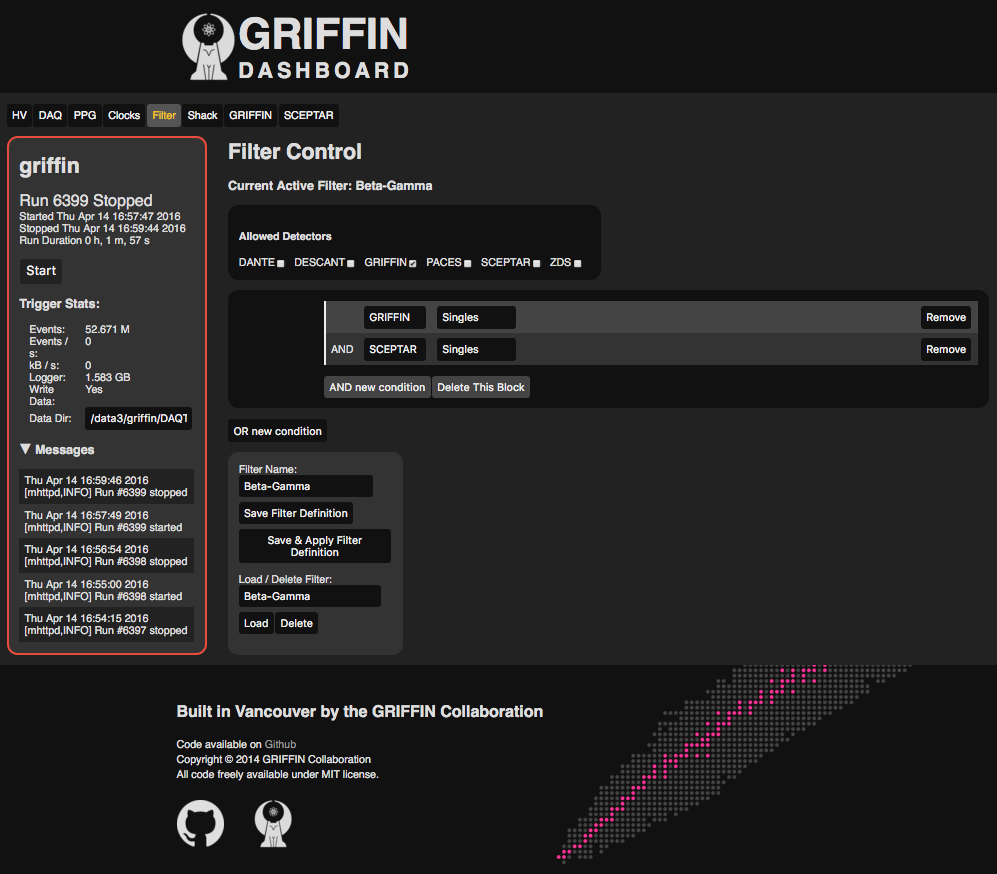}
\caption{Screenshot of the Dashboard User Interface showing the control page of the master filter.}
\label{fig:User_Dashboard_Filter}
\end{figure}

A user-friendly interface has been developed to interact with the GRIFFIN DAQ system in order to make regular tasks such as rate-monitoring or parameter changes straight-forward to complete. The user interface is integrated into the MIDAS DAQ system in order to read and write parameters from a single, centralized location called the Online DataBase (ODB). At the beginning of a data collection period the parameters are read from the ODB and written to all electronics modules to ensure synchronized values. A copy of the entire ODB is written to the disk file at the start and end of each data collection period for reference during analysis. All user interface pages are provided in web browsers and various pages are provided from two different servers. The pages are constructed primarily using Javascript. The mhttpd server of MIDAS provides a general Dashboard interface page which includes control of the main parameters of the GRIFFIN facility such as cycle control or master filter logic (shown in Figure \ref{fig:User_Dashboard_Filter}). In addition, each GRIF-16 digitizer runs a separate server to provide an interface page for that individual digitizer. When a parameter is changed from a page running from an individual digitizer the value is immediately synchronized in the ODB and percolated to all other interface pages.

\begin{figure}
\centering
\fbox{\includegraphics[width=1.0\linewidth]{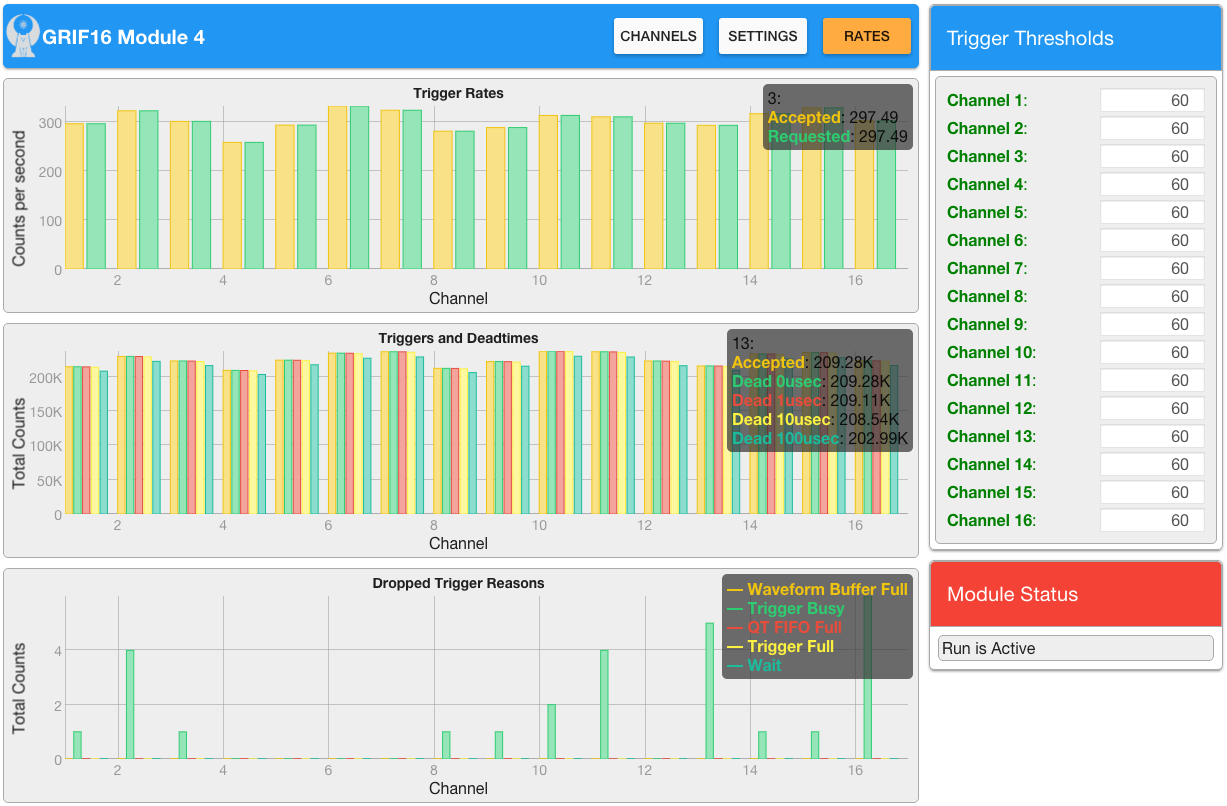}}
\caption{Screenshot of an example GRIF-16 digitizer User Interface showing the rates, scalers and any error reporting for the 16 channels of the module.}
\label{fig:User_GRIF16_Rates}
\end{figure}

\begin{figure}
\centering
\fbox{\includegraphics[width=1.0\linewidth]{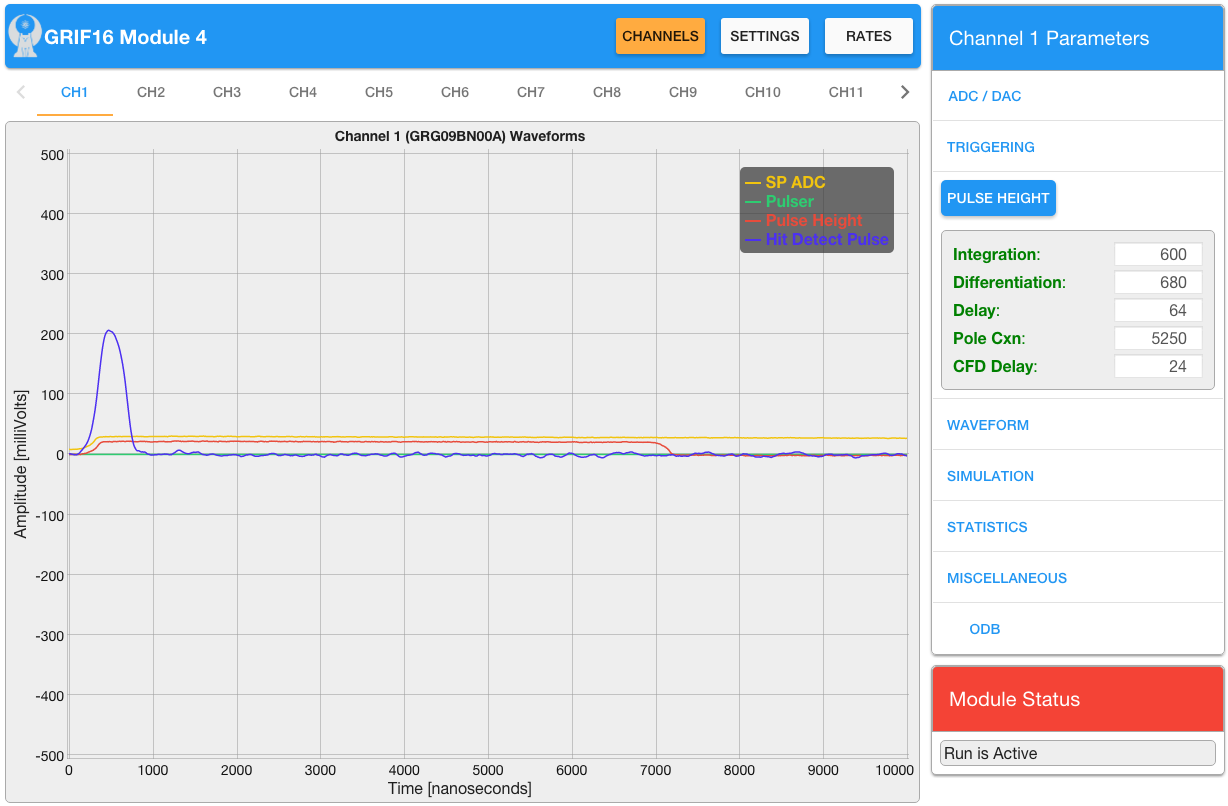}}
\caption{ Screenshot of an example GRIF-16 digitizer User Interface for a single channel. The waveforms update in real time similar to a digital oscilloscope. The menus on the right allow the user to change parameters of the signal processing algorithms.}
\label{fig:User_GRIF16_Chan}
\end{figure}

A screenshot of an example GRIF-16 digitizer User Interface showing the rates, scalers and any error reporting for the 16 channels of the module is shown in Figure \ref{fig:User_GRIF16_Rates}. A screenshot of an example GRIF-16 digitizer User Interface for a single channel is shown in Figure \ref{fig:User_GRIF16_Chan}. The waveforms update in real time similar to a digital oscilloscope. The menus on the right allow the user to change parameters of the signal processing algorithms during setup mode and when data is not being collected. A software lock prevents parameters being changed during data collection as this would create difficult analysis situations. 

\section{Performance of the System}
\label{sec:Performance}

The majority of the experiments performed to date with GRIFFIN have been focused on measurements relevant to nuclear structure and nuclear astrophysics investigations. In such measurements the performance features of primary relevance are detection efficiency and spectral quality. A main design goals of the GRIFFIN DAQ system was to ensure these features up to high counting rates in order to collect ultra-high statistics datasets and search for very low-intensity decay branches. The performance of the GRIFFIN DAQ system is presented here.

\subsection{Low-Energy Thresholds}
\label{sec:Thresholds}

\begin{figure}
\centering
\includegraphics[width=1.0\linewidth]{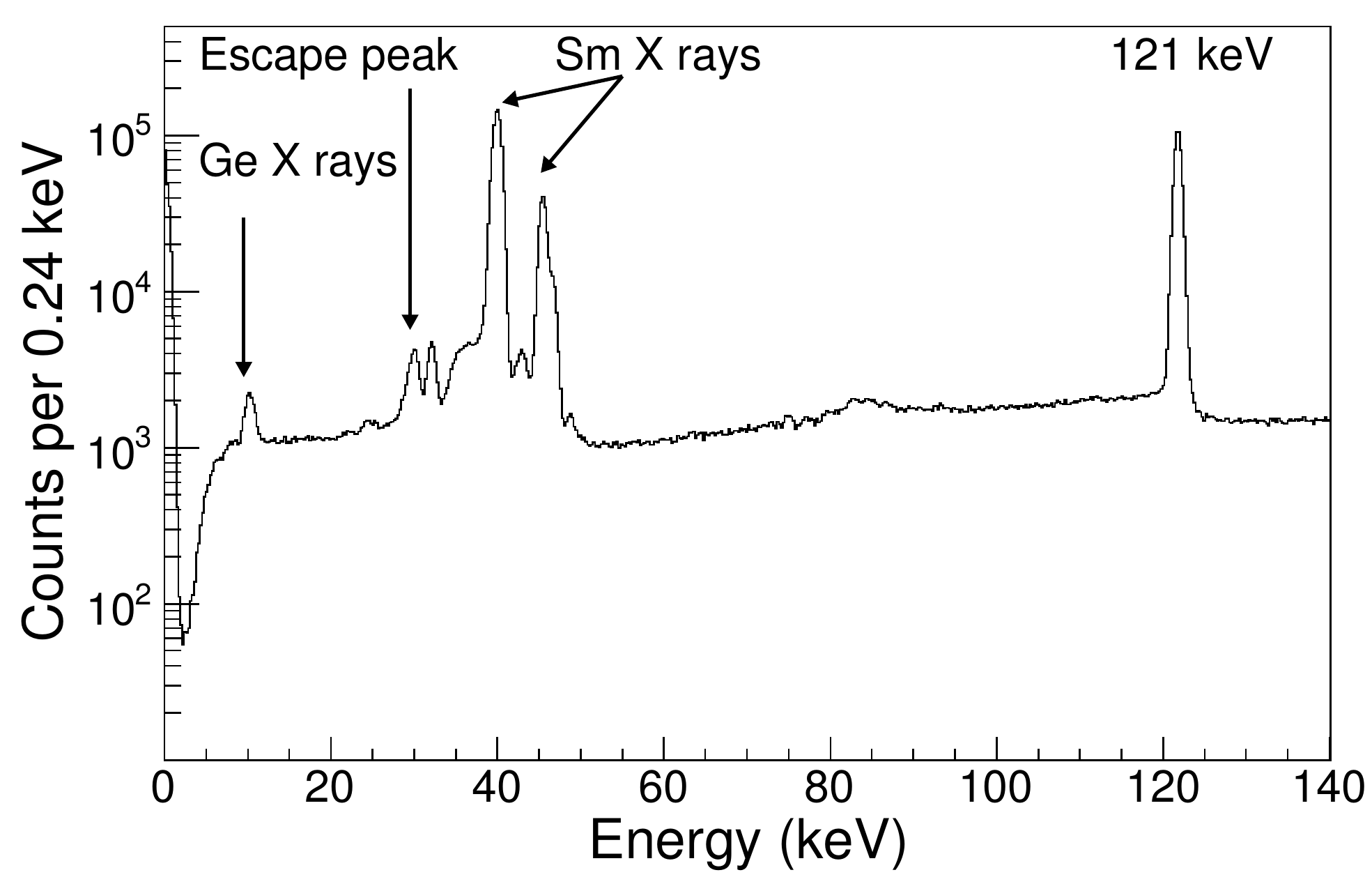}
\caption{HPGe energy spectrum of a $^{152}$Eu source collected with a GRIF-16 digitizer showing the low-energy region. The energy resolution of the 122~keV peak is 1.07~keV. The escape peak at 30~keV results when an incident Sm X ray is detected but the Ge X ray escapes the crystal. The Ge X ray peak is visible at 9.9~keV.}
\label{fig:LowThres}
\end{figure}

The hit detection algorithm in the GRIF-16 (Section \ref{sec:SignalProcessHit}) successfully triggers on 5~keV equivalent simulated pulses injected into the ADC data of a real detector baseline. When examining real GRIFFIN HPGe detector signals, as can be seen in Figure \ref{fig:LowThres}, the germanium K x-ray peak at 9.9~keV can clearly be identified which is the result of photons moving between crystals inside the HPGe clover crystal housing. These events can be used in the add-back method to improve the full-photopeak detection efficiency \cite{Rizwan16}. However, with a thickness of 1.5~mm the aluminum housing around each clover detector heavily attenuates photons of this energy from reaching the crystals from outside. Therefore the DAQ is capable of triggering at a lower energy than the HPGe detectors are expected to observe from external sources.

\subsection{Energy Resolution}
\label{sec:Energy_Res}

The Full-Width at Half Maximum (FWHM) of the full-energy photopeak for the 122~keV and 1332~keV gamma rays from $^{152}$Eu and $^{60}$Co detected in a single GRIFFIN HPGe crystal are typically 1.07(1) and 1.87(1)~keV respectively when measured using a GRIF-16 digitizer at low counting rate.

\begin{figure}
\centering
\includegraphics[width=1.0\linewidth]{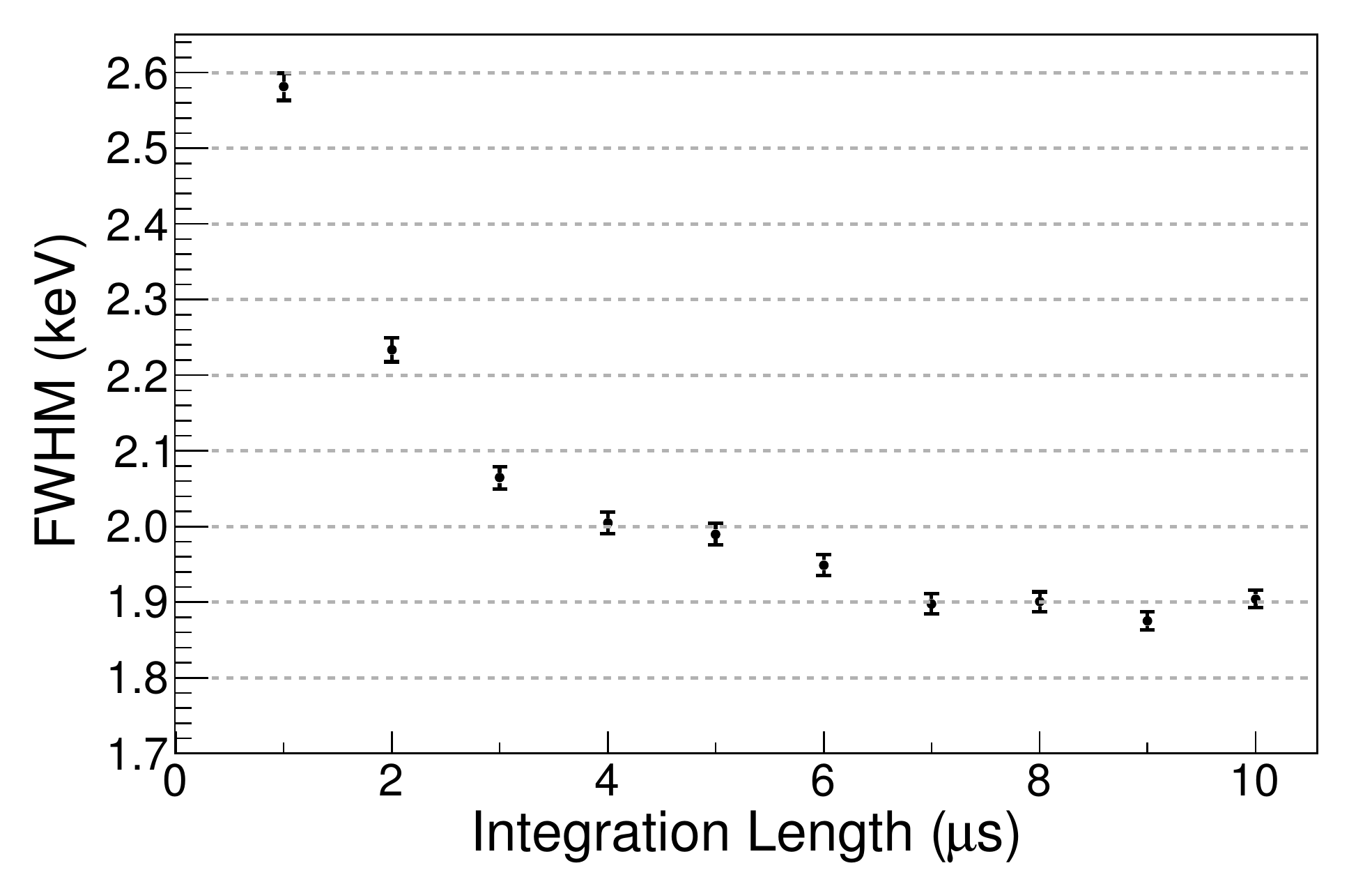}
\caption{FWHM resolution (keV) of the 1332~keV gamma ray from $^{60}$Co plotted as a function of the integration length ($\mu$s) used in the pulse-height evaluation algorithm. Data are for a single HPGe crystal counting at low rate ($\approx$ 300 Hz).}
\label{fig:fwhm_int}
\end{figure}

The energy resolution as a function of the integration time, $T^E_{int}$, is shown in Figure \ref{fig:fwhm_int}. In each case the differentiation time, $T^E_{dif}$, was 1~$\mu$s longer than $T^E_{int}$. The energy resolution improves with increasing integration time and is essentially constant above 6~$\mu$s. The reduction in performance with shortening integration time is a limiting factor in the energy resolution of pile-up events at high counting rates, as discussed in Sections \ref{sec:Pile-up} and \ref{sec:HighRateSpectralQuality}.

The linearity of the energy response in the GRIF-16 digitizer has been measured in the energy range from 122 to 3253~keV using radioactive sources of $^{152}$Eu and $^{56}$Co. The linearity in gain was found to be better than 0.3~keV in this energy range in comparison to 0.2 keV measured for the GRIFFIN HPGe detectors with a TIG-10 digitizer \cite{Rizwan16}.

When counting at high rates the DC offset of the baseline of a HPGe signal will vary considerably as seen in Figure \ref{fig:Energy_3pile-ups}a. This means different regions of the ADC will be used to measure the same energy for different hits. The linearity was also investigated for different regions of the ADC by examining the energy response to the 59.5~keV gamma ray from a $^{241}$Am source  at $<$1kHz and adjusting the DC offset manually using the input DAQ of the GRIF-16 module. The linearity of the response was also found to be better than 0.3~keV over the full ADC range.

\subsection{Coincident-Timing Resolution}
\label{sec:Timing}

\begin{figure}
\centering
\includegraphics[width=1.0\linewidth]{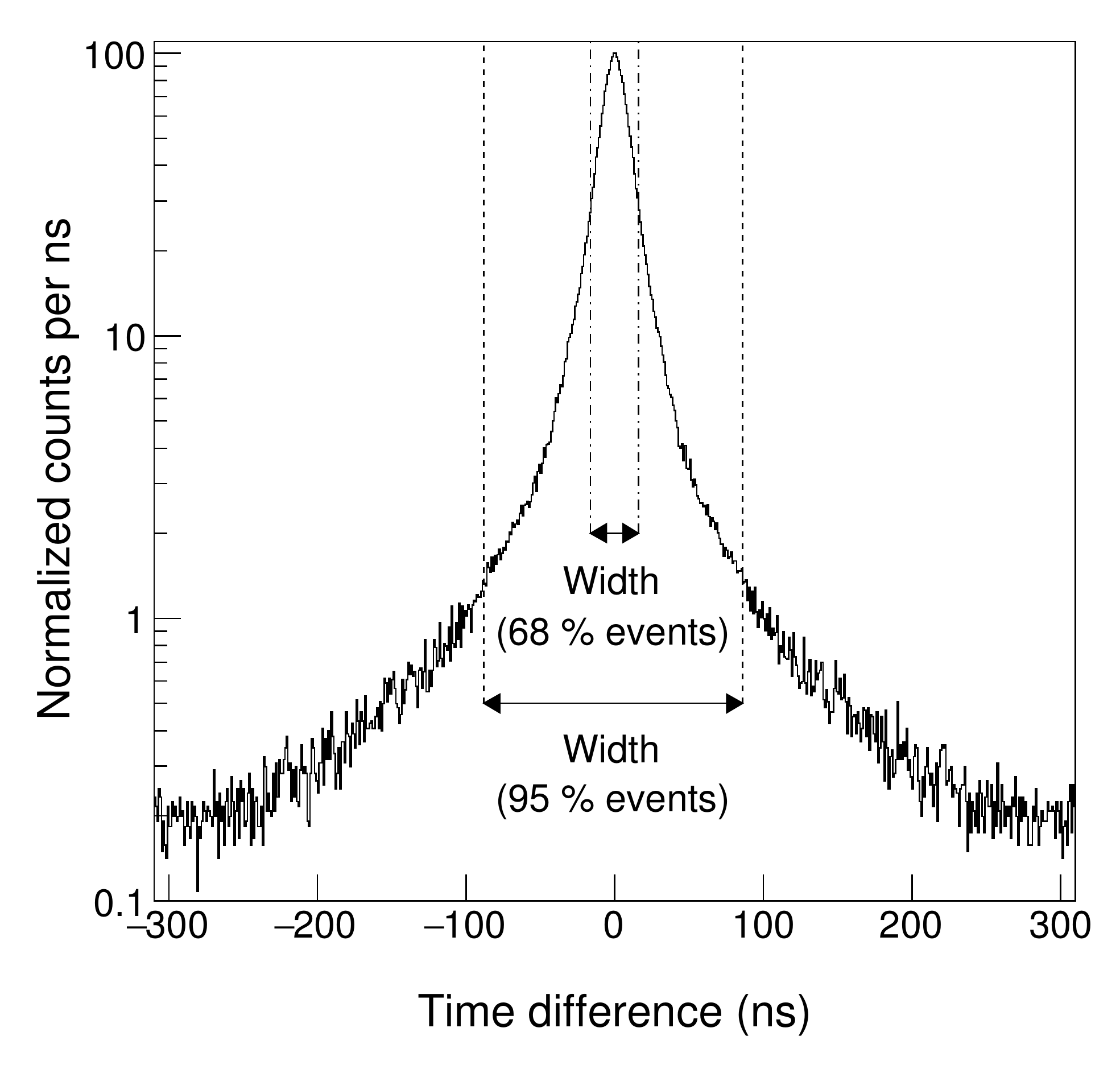}
\caption{Time difference between coincident gamma-gamma events from GRIFFIN HPGe signals processed in the GRIF-16 digitizer by the timing algorithm. A threshold of 20~keV has been applied to both coincident energies. An additional requirement of single interactions (no pile-up) within each HPGe crystal was imposed.
At an incident rate of $<$5~kHz in each crystal, 68\% of the coincidence events have a time difference less than 32(2)~ns and 95\% of the coincidence events are contained within a width of 174(8)~ns.}
\label{fig:CFD}
\end{figure}

The result of the timing algorithm is shown in Figure \ref{fig:CFD} where the time difference between coincident gamma-gamma events from a $^{60}$Co source are processed in the GRIF-16 digitizer. A condition that the crystal multiplicity is equal to two is imposed and the energy required is greater than 20~keV. The shape of the distribution is not a Gaussian so the usual FWHM cannot be quoted. The width of this coincidence timing peak containing 68~\% of events is 32(2)~ns and 95\% of the data is within 174(8)~ns.

\begin{figure}
\centering
\includegraphics[width=1.0\linewidth]{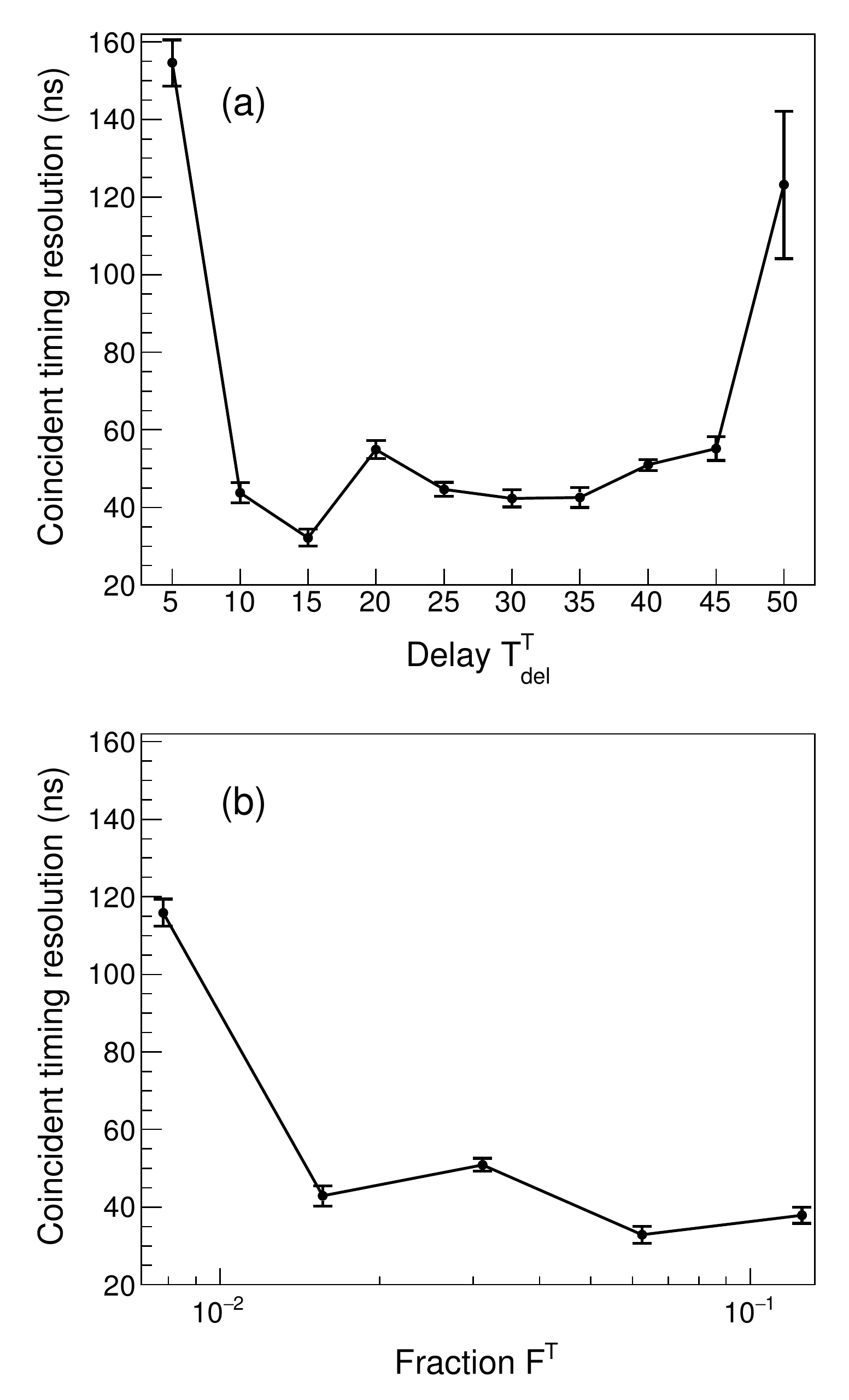}
\caption{The width containing 68~\% of the time difference peak between coincident gamma-gamma events from GRIFFIN HPGe signals as a function of the Delay (panel a) and Fraction (panel b) parameter values used in the timing algorithm. The Delay parameter was maintained at a constant value (equal to 14) during each Fraction setting in panel b. The time difference between all events with energy $>$20~keV from a $^{60}$Co source is included.}
\label{fig:CFD_syst}
\end{figure}

The performance of the timing algorithm is sensitive to the values of the Delay, $T^T_D$, and Fraction, $F^T$, parameters. The systematic behavior can be seen in Figure \ref{fig:CFD_syst}. The time difference between all events from HPGe signals with energy $>$20~keV viewing a $^{60}$Co source is included in this analysis.

\subsection{Spectral Quality at high counting rate}
\label{sec:HighRateSpectralQuality}

\begin{figure*}
\centering
\includegraphics[width=0.90\textwidth]{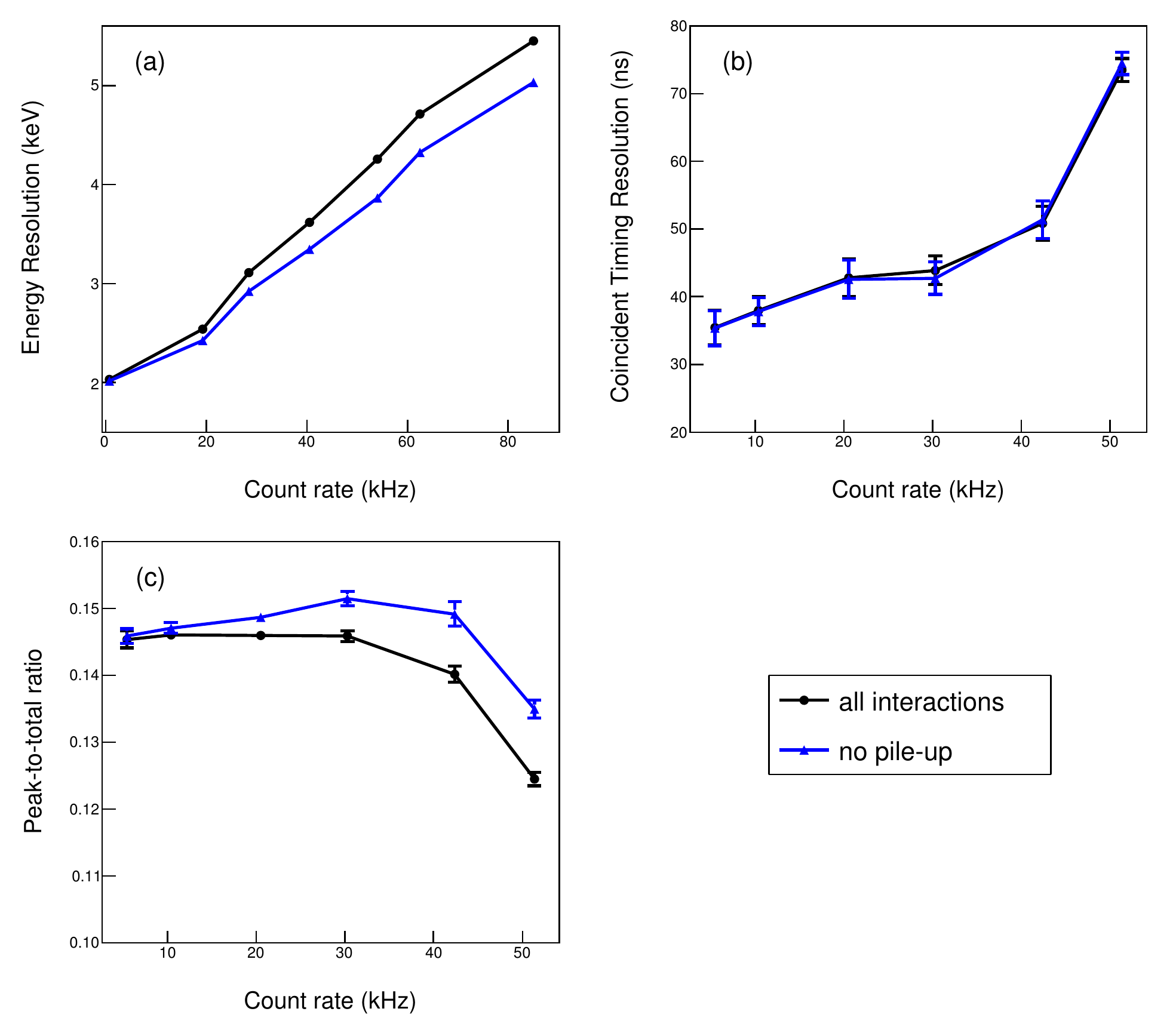}
\caption{Energy resolution (a), coincidence-timing resolution (b) and peak-to-total ratios (c) as a function of individual crystal counting rate collected using the GRIF-16 digitizer and GRIFFIN clovers observing a $^{60}$Co source. See text for details.}
\label{fig:SixPanel}
\end{figure*}

The systematic behavior of various performance metrics is shown in Figure \ref{fig:SixPanel} as a function of individual HPGe crystal counting rate. Ideally the system will be able to collect data at high counting rates with the same spectral quality as at low counting rates. In order to achieve a good spectral quality the energy resolution and peak-to-total ratio must remain high as a function of counting rate. The coincident-timing resolution at all rates is important for the accurate subtraction of random-coincidence background contributions. In addition the relative efficiency of the detector system must also remain high with increasing rate which is discussed in Section \ref{sec:HighRateEfficiency}.

The data for the energy resolution, timing resolution and peak-to-total plots were collected using the GRIF-16 digitizer and GRIFFIN clovers observing a $^{60}$Co source. The crystal counting rate was controlled by changing the source-to-detector distance. Note that this also changes the solid angle subtended by the detector and therefore the probability of true-coincidence summing is changed at the same time as the counting rate. This does not affect the performance results obtained for the aforementioned metrics. 

Note that in all examples of ``all interactions'' and ``no pile-up'' that these conditions are applied in the offline analysis with all data being recorded to disk by the DAQ system.

The pile-up recovery in the GRIFFIN DAQ system does come at a cost to energy resolution as can be seen in Figure \ref{fig:SixPanel}a in the comparison between single-interaction and pile-up events, but is essential to maintain a high detection efficiency as discussed in Section \ref{sec:HighRateEfficiency}. There is an increase in energy resolution as a function of counting rate that is almost linear and following the trend of the pile-up fraction (Figure \ref{fig:PileupFraction}). This increase is in part due to the unavoidable reduction in the integration period, $T^E_{int}$, applied during the pulse-height evaluation of pile-up events. The energy resolution increases from 2.0keV at $\sim$1~kHz to $>$4~keV above 50~kHz. This will obviously reduce the resolving power of the spectrometer and limit the ability to detect very weak intensity decays. It may be possible to further improve the energy resolution algorithm performance at high counting rates.

One of the primary objectives in high-statistics experiments is the construction of level schemes through a gamma-gamma coincidence analysis. The resolution of coincidence timing is important in order to reject false and background coincident events from the spectra. Figure \ref{fig:SixPanel}c shows that the performance of the timing algorithm remains the same to count rates of 40~kHz but worsens at higher rates.

The peak-to-total ratio remains approximately constant up to counting rates of 40~kHz (Figure \ref{fig:SixPanel}c) showing that the spectral shape remains very similar.

\subsection{Relative Efficiency at high counting rate}
\label{sec:HighRateEfficiency}

Figure \ref{fig:PileupFraction} shows that it is essential to handle pile-up events in an effective way in order to maintain a good detection efficiency because at counting rates of $\sim$50~kHz around 45\% of events pile-up within a 6~$\mu$s differentiation period. If one simply rejects pile-up events by considering only single-interaction events then the relative detection efficiency is greatly reduced at high counting rates. In the GRIFFIN DAQ system however, the majority of pile-up events are recovered and therefore a high relative detection efficiency is maintained. Very little data loss is of course also a requirement when counting up to 50~kHz per crystal which is equivalent to roughly 134~MB per second for 64 HPGe crystals.

\begin{figure}
\centering
\includegraphics[width=1.0\linewidth]{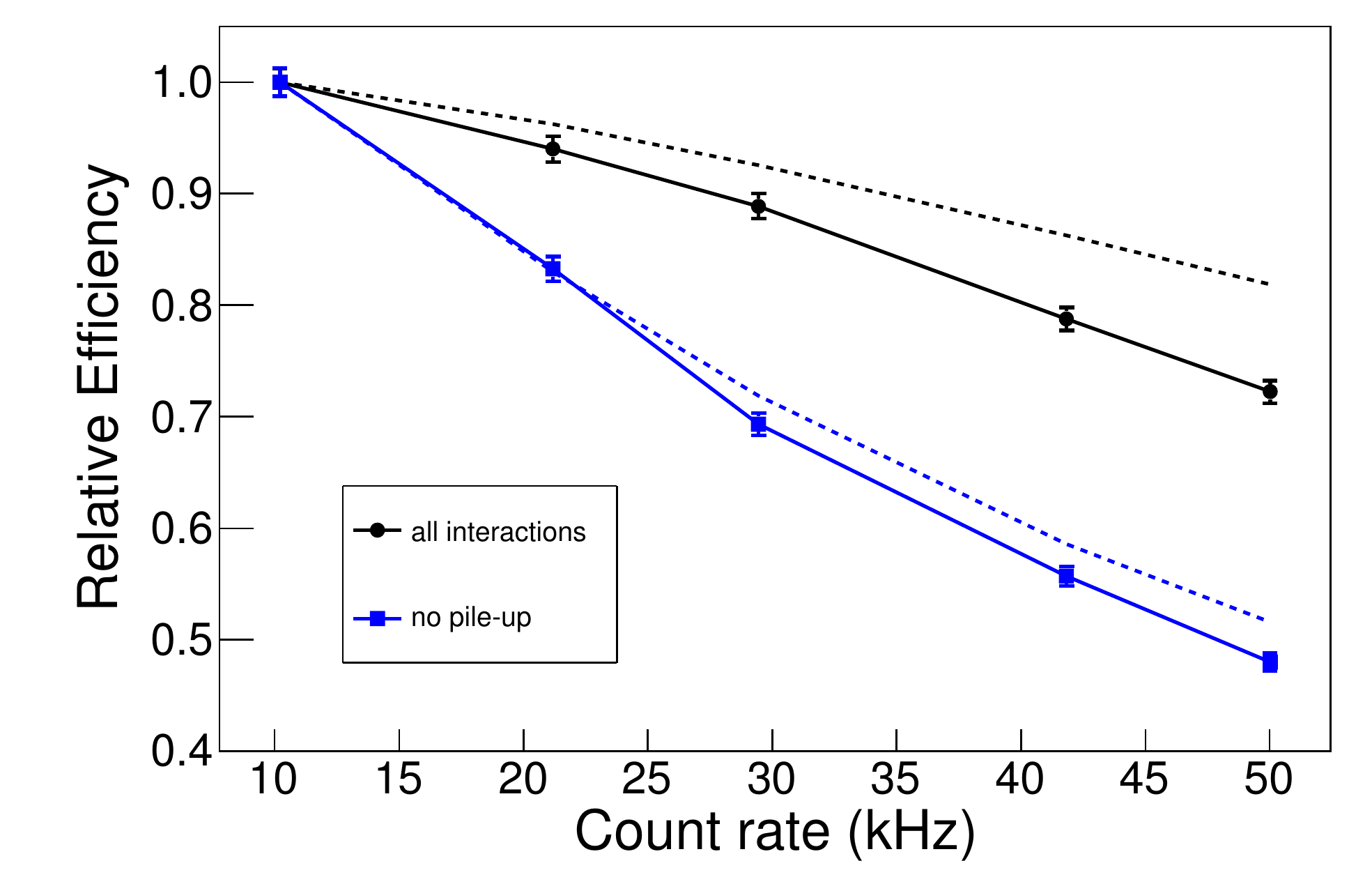}
\caption{Relative efficiency at 1173~keV as a function of individual crystal counting rate collected using the GRIF-16 digitizer and GRIFFIN clovers observing a $^{60}$Co source. Calculated relative efficiency curves considering all interactions and single hits are shown as dotted lines. See text for details.}
\label{fig:RelEff}
\end{figure}

Figure \ref{fig:RelEff} shows the relative efficiency as a function of crystal counting rate. In the determination of the relative efficiency, the probability of true-coincidence summing is an important consideration and therefore the distance of the source must be consistent for the measurement at each counting rate. In this case a $^{60}$Co source was placed at a fixed distance from the detector and data collected for 3~mins at each counting rate. The counting rate was controlled by changing the distance of a $^{137}$Cs source with respect to the detector. Here the parameters $T^H_{DT}=1.2\mu$s and $T^E_{diff}=8\mu$s. The data points with solid lines represent the experimental results and the dashed lines are the result of the equations discussed in Section \ref{sec:Pile-up} and \ref{sec:Append_summing_pile-up}.

\begin{figure*}
\centering
\includegraphics[width=0.9\linewidth]{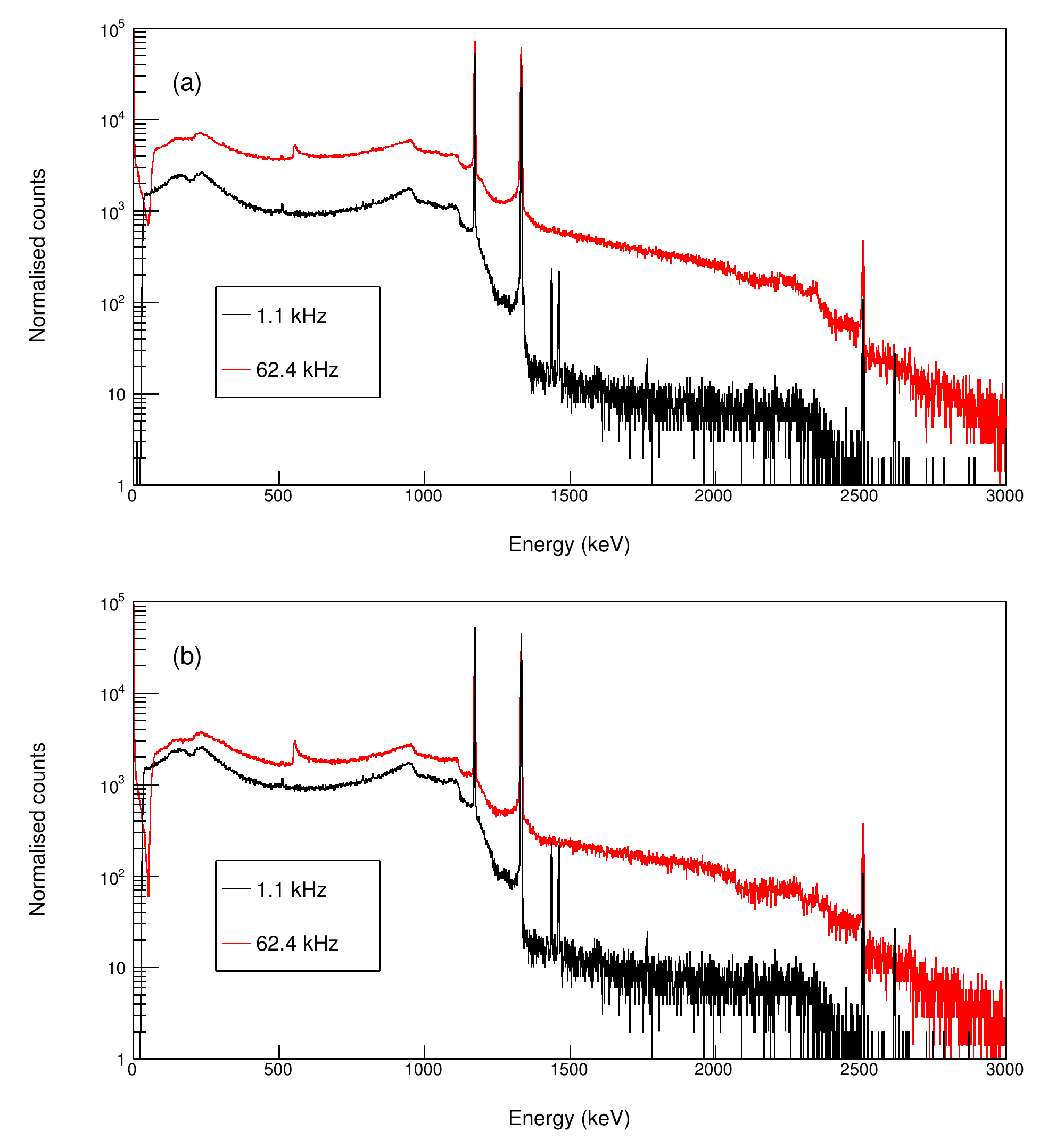}
\caption{GRIFFIN energy spectra for (a) only single interactions, and (b) all interactions, collected using a $^{60}$Co source for a single HPGe channel at counting rates of 1 and 62~kHz. The spectra are normalized according to the number of counts in the 1173~keV photopeak, corrected for the probability of detecting the full-energy. The feature at $\sim$600~keV in the high-rate spectrum is the result of Compton scattering from a piece of Pb shielding not present in the low-rate data. A number of peaks from room background can be seen in the low-rate spectrum including the radioactivity of $^{138}$La present in the LaBr$_3$ detectors.}
\label{fig:highrate_energy}
\end{figure*}

The apparent relative efficiency at higher counting rates will be reduced by random-coincidence summing and by pile-up scenarios that are not deconvolved in the data analysis process.

A comparison of the gamma-ray energy spectra from a $^{60}$Co source collected at crystal counting rates of 1 and 50~kHz is shown in Figure \ref{fig:highrate_energy}. In each case the source-to-detector distance was $\sim$50~cm. The spectra are normalized by the number of counts in the 1332~keV peak. The upper panel shows only single interactions equivalent to pile-up rejection whereas the lower panel shows all interactions, the equivalent of pile-up recovery. In this example, at the high counting rate the same level of statistics can be collected in $<$3\% of the data collection period required at the low counting rate. Good spectral quality is seen in both spectra indicating that the design goal for high counting rate operation has been achieved.

\subsection{Fixed-Deadtime Scalers}
\label{sec:Perform_Scalers}

The functionality of the fixed-deadtime scalers is described in section \ref{sec:Scalers}. The fixed deadtimes imposed on each scaler were measured independently using a variation of the paired-source method introduced by A. P. Baerg \cite{Baerg65} where one of a pair of random sources is replaced with a periodic pulse of known frequency. Provided that the pulse period is greater than the non-extending deadtime applied to a given channel, the deadtime may be measured by comparing the combined counting rate of the source and periodic pulse with the rate of the source alone. The measured deadtime $\tau_{m}^{d}$ is given by

\begin{equation}
\tau_{m}^{d} = \frac{1}{R_{s}}\sqrt{1-\frac{R_{c}-R_{s}}{R_{p}}},
\label{eq:dt}  
\end{equation}    

\noindent where $R_{s}$ is the random source rate, $R_{p}$ is the frequency of the periodic pulse and $R_{c}$ is the combined rate of the source and periodic pulse measured by the scaler. The measurement was performed as follows. A 9.28 $\mu$Ci $^{137}$Cs source was placed near the center of a single hemisphere of the GRIFFIN HPGe array. The periodic square-wave output of an Agilent 33522A arbitrary waveform generator was used to produce a logic pulse from an ORTEC GG8020 gate generator that was connected to the test input connector of 9 HPGe crystal in the same hemisphere. As the Pulser signal was delivered directly to the preamplifier of each crystal, both periodic and random pulses from the $^{137}$Cs decay were collected and processed together by a GRIF-16 for a given channel.

\begin{figure}
\centering
\includegraphics[width=1.0\linewidth]{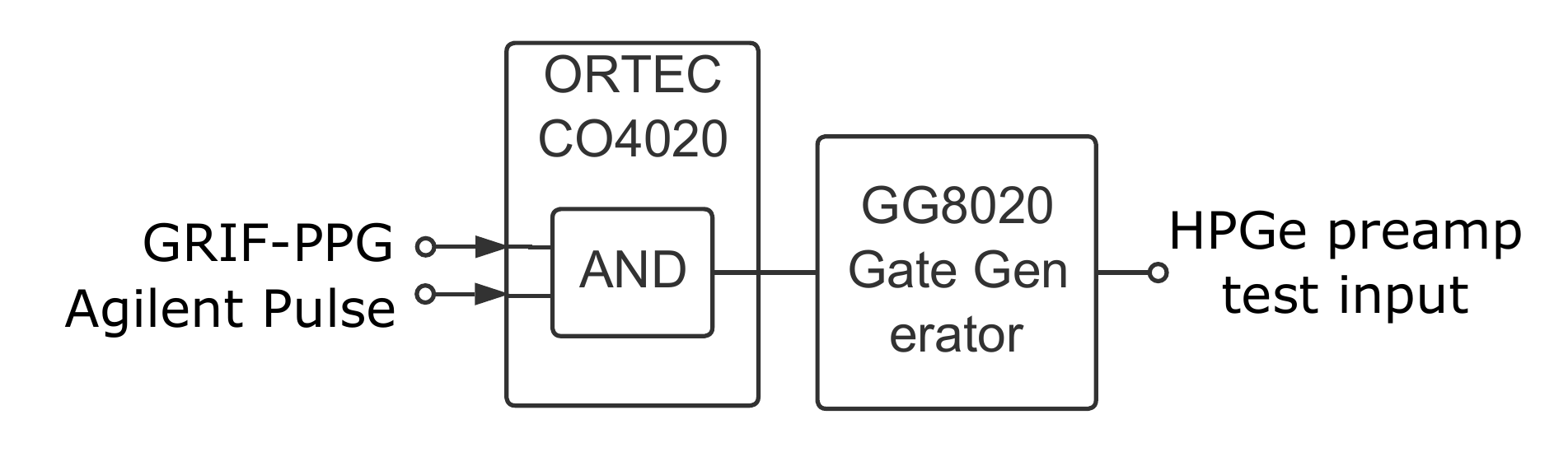}
\caption{Schematic of the experimental setup used to measure the fixed deadtime of the precision deadtime scalers.}
\label{fig:setup}
\end{figure}

The GRIF-PPG module was used in conjunction with analogue electronics to control the measurement cycle, comprised of 10 seconds of measurement with the Pulser `on' ($R_{c}$) and 10 seconds with the Pulser `off' ($R_{s}$). In this way no cables were unplugged or connected when changing between ($R_{c}$) and ($R_{s}$). The first and last two precision deadtime scaler readings obtained during transitions between the two states were omitted from the analysis. A diagram of the test setup is shown in Figure \ref{fig:setup}.
The AND signal of the GRIF-PPG and Pulser generated by an ORTEC CO4020 logic unit was used as input to a ORTEC GG8020 Gate Generator. The Gate Generator was used to shape the width of the NIM pulser signal ($\approx$ 3 $\mu$s) to ensure no modification of the signal processing algorithm in the GRIF-16 was necessary (an unmodified square-wave pulse from the Agilent pulser is too large and distorts the HPGe preamplifier output signal). A total of four measurements were conducted, each using a different Pulser frequency, where $R_{p}$ was varied from 2-20 kHz. $R_{s}$ remained $\approx$3~kHz per HPGe channel. Each measurement lasted $\approx$90~cycles.

\begin{center} 
\begin{table}
\caption{Summary of deadtime measurements performed for a single GRIFFIN HPGe channel. $\tau_{i}^{d}$ is the non-extending deadtime imposed on each fixed-deadtime scaler channel and $\tau_{m}^{d}$ is the deadtime measured according to Eq. \ref{eq:dt}. $R_{s,c}$ are the mean counting rates of the source and of the source plus Pulser respectively, averaged over all measurement cycles. $R_{p}$ is the Pulser frequency.}
\begin{tabular}{ccccc}
\hline
$R_{s}$&$R_{p}$&$R_{c}$&$\tau ^{d}_{i}$&$\tau ^{d}_{m}$\\
kHz&kHz&kHz&$\mu$s&$\mu$s\\[2mm]
\hline
2.425(47)&20&22.26$_{14}^{13}$&0&1.7$_{1.7}^{1.3}$\\[2mm]
2.421(47)&20&22.23$_{14}^{13}$&1&2.0$_{2.0}^{1.0}$\\[2mm]
2.370(46)&20&21.44$_{14}^{11}$&10&9.9$_{1.4}^{1.1}$\\[2mm]
1.951(35)&2&3.247$_{55}^{57}$&100&100.0$_{9.8}^{9.7}$\\[2mm] 
\hline
\end{tabular}
\label{tab:table1}
\end{table}
\end{center}

\begin{figure}
\centering
\includegraphics[width=1.0\linewidth]{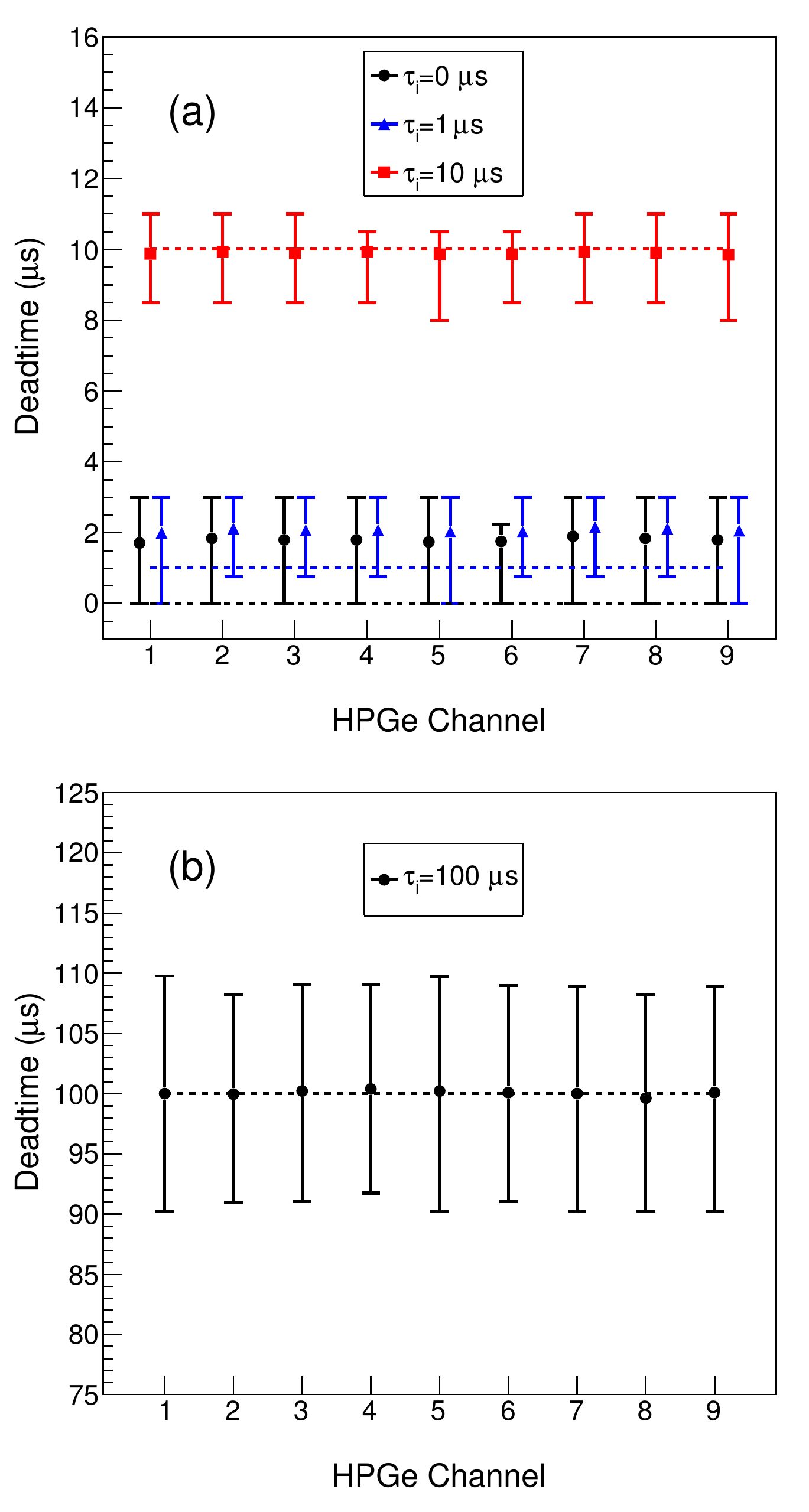}
\caption{Deadtime determined using Equation \ref{eq:dt} plotted against GRIFFIN HPGe channel number for (a) imposed deadtimes $\tau_{i}$ of 0, 1 and 10 $\mu$s and (b) 100 $\mu$s. Data obtained for $\tau_{i}$ $\leq$ 10 $\mu$s used a Pulser frequency of 20 kHz. Due to a limiting factor of the paired-source method where the arrival time of the periodic pulses must be greater than the imposed deadtime, measurements with $\tau_{i}$ = 100 $\mu$s were performed using a 2 kHz Pulser frequency.}
\label{fig:deadtime}
\end{figure}

Results obtained for a single GRIFFIN HPGe channel are shown in Table \ref{tab:table1}. The results from all nine channels tested are plotted in Figure \ref{fig:deadtime}. For imposed deadtimes of 0,1 and 10 $\mu$s, the data correspond to tests performed using a 20,000.000(20)~kHz Pulser rate. For an imposed deadtime of 100~$\mu$s, a 2,000.000(2)~kHz Pulser rate (2~ms pulse period) ensured that the arrival time between pulses was greater than the imposed deadtime.

A combination of statistical analyses taking into account both the non-random nature of the periodic pulses and the correlation between the uncertainties in $R_{s}$ and $R_{c}$ were used to determine the final uncertainty in $\tau_{m}^{d}$ \cite{Barlow06}. $\tau_{m}^{d}$ and $\tau_{i}^{d}$ agree for all channels within uncertainty limits, typically within $\pm$10 $\%$ for $\tau_{i}^{d}$=10,100 $\mu$s, with larger relative uncertainties observed for smaller imposed deadtimes.

\section{Precision Measurements}
\label{sec:PrecMeas}

A thorough understanding of detector and electronics pulse pile-up effects are especially important when making precision measurements \cite{Grinyer07}. In addition it is desirable for the deadtime of each detector channel to be determined with good accuracy. The GRIFFIN system provides the information in the data required to assess these factors.

The precision deadtime scalers discussed in sections \ref{sec:Scalers} and \ref{sec:Perform_Scalers} are not suitable for precision half-life measurements using HPGe detectors because of the relatively low peak-to-background ratio in gamma-ray spectra. The use of PMT signals with such scalers can also be unreliable. However this firmware could be employed with a detector such as the 4$\pi$ continuous gas-flow proportional $\beta$ counter described in Ref. \cite{Ball14} to make precision half-live measurements. This application will be looked at in the future.

The performance of the system for making precision half-life measurements remains to be demonstrated but each of the required features are already implemented in the firmware and described in the earlier sections.

\section{Future Development Plans}
\label{sec:Future}

The system that has been implemented to date uses 5~Gb/s links between digitizer and collector modules, as well as from the GRIF-C Master module to the backend computer running MIDAS. A 5~Gb/s link is more than adequate for the $\sim$34MB/s of data produced from 16 channels of HPGe counting at 50~kHz each, or for associated ancillary detectors. This link capacity is also sufficient to connect the GRIF-C Slave modules to the GRIF-C Master module as long as a sufficient number are used. The master filter algorithm can process 1 event per clock which is equivalent to 100 million events per second or $\sim$35~Gb/s.
The present single 1~Gb/s link from the GRIF-C Master to the MIDAS system is a current limitation, throttling the overall DAQ through-put to $<$125MB/s. This implementation has been used to demonstrate the performance for a subset of the GRIFFIN array up to 50~kHz but limits global counting rate for the full detector system. At the backend, the MIDAS system has been demonstrated to operate at more than 700~MB/s using 10~Gb/s network connections and a disk server.

It is planned to implement a 10~Gb/s link connection from the GRIF-C Master module to the backend computer in order to relieve the main through-put limit of the present system. The Serial RapidIO architecture \cite{RapidIO} for high-performance packet-switched, interconnection between all devices in the system will also be implemented, more for standardization and firmware portability rather than performance gain. The hardware was designed with these capabilities in mind and the necessary firmware development will be completed in the near future.

A fast-sampling FMC card is under development which will use the GRIF-16 module as a carrier board. This card is based around the dual-channel, 12-bit AD9234 ADC chip which will enable either four channels of 500~MHz sampling, or two channels of 1~GHz sampling. This card is at the prototype stage and will be used to process the signals directly from the PMT output of scintillators.

A 32-channel, 65~MHz sampling FMC card is under development which will use the GRIF-16 module as a carrier. This is a cost-effective solution using two AD9249 ADC chips for integrating a highly-segmented auxiliary detector system, such as a silicon barrel array similar to the SHARC detector \cite{Diget11}.

\section{Summary}
\label{sec:summary}

A new digital data acquisition system has been commissioned at TRIUMF-ISAC to collect data with the GRIFFIN HPGe facility. The new system is comprised of custom designed and built electronics modules. The aspects of the system suitable for performing high-precision measurements relevant to the study of Fermi superallowed beta decays are discussed. The capabilities of the system to collect data of good spectroscopic quality at counting rates up to 50~kHz per HPGe crystal has been demonstrated.

\section{Acknowledgments}

N. Bernier, J. Measures and J. Pore are thanked for their assistance in collecting data used to construct versions of some figures in this manuscript. A.B.G. is grateful for support from the Department of Nuclear Physics, Australian National University during the preparation of this manuscript. C.E.S. acknowledges support from the Canada Research Chairs program. This work was supported by the Natural Sciences and Engineering Research Council of Canada. The GRIFFIN infrastructure has been funded jointly by the Canada Foundation for Innovation, TRIUMF and the University of Guelph. TRIUMF receives funding via a contribution agreement through the National Research Council Canada.

\appendix
\section{The probability of random-coincidence-summing and pile-up}
\label{sec:Append_summing_pile-up}

If one assume a Poisson distributed process of rate $R$, then the probability of there being no hits in the following time period $\Delta t$ is
\begin{equation}
P(0) = e^{-R\Delta t}.
\label{Append_eq.1}
\end{equation}
The probability that there \emph{is} a hit in the time period $dt$ is:
\begin{equation}
R(1) = R dt.
\label{Append_eq.2}
\end{equation}
The variable $t$ must be integrated over a period of time to obtain a physical probability.
These two equations can be used to calculate the probability of different event types in the GRIFFIN DAQ system. We define an event as a grouping of occurrences that are within a time period $T_{PP}^E$ of each other. 
The first gamma-ray interaction, here labeled as hit 0 at $t_0$, opens up two windows: a fixed dead time of duration $T_{DT}^H$ and a pulse processing window of duration $T_{PP}^E$. The pulse processing window is equal to the $T^E_{diff}$ plus the risetime of the pulse. This is not a fixed time but the risetime is only a small addition to the $T^E_{diff}$. Any hits that occur before the end of the dead time window are summed and not identified by the system as separate gamma-rays. Any interactions that occur after the end of the dead time window but before the end of the pulse processing window opens a new pair of windows and extends the possible time for other interactions to be grouped into this event. In addition, in order to be the first hit, there must be no other hits in a time window $T_{PP}^E$ immediately preceding the initial hit.

In order for a hit to be a ``single interaction" event, no hits should occur within the time period $T_{PP}^E$ on either side of $t_0$. Therefore, using Equation \ref{Append_eq.1}, the probability that hit 0 is in a single interaction event is:
\begin{align}
P &= e^{-2 R T_{PP}^E}
\end{align}

When there is a second hit within the pulse processing window then it can either be summed or piled-up. The probability of the two hits summing is equal to,
\begin{align}
P &= e^{-2RT_{PP}^E}R\int_{t_1=0}^{T_{DT}^E}dt_1\\
&= e^{-2RT_{PP}^E}RT_{DT}^E
\end{align}
and the probability of the two hits being piled-up is,
\begin{align}
P &= e^{-2RT_{PP}^E}R\int_{t_1=T_{DT}^H}^{T_{PP}^H}e^{-Rt_1}dt_1\\
&= e^{-2RT_{PP}^E}R\left [- \frac{1}{R}e^{-Rt_1} \right ]_{t_1=T_{DT}^H}^{T_{PP}^H}\\
&= e^{-2RT_{PP}^E}(e^{-R T_{DT}^H}-e^{-R T_{PP}^E})
\end{align}.

The general expansion of this to higher numbers of hits must take into account the probability of summing occurring between any pair of hits in the full pile-up event and is determined as
\begin{equation}
P=e^{-R T_{DT}^H}\prod_{i=1}^{N}\left (dt_N \right )e^{-R t_e}
\label{eq.general}
\end{equation}
where $N$ is the number of hits and $t_e$ is either $T^H_{DT}$ or $T^E_{PP}$ depending on if it is a summing or pile-up scenario. Table \ref{tab:Append_sum_equations} lists the equations for each scenario involving  up to four hits.

\begin{table*}[ht]
   \begin{center}
   \caption{\label{tab:Append_sum_equations}Probability calculations for the probability that a single gamma-ray interaction is the gamma-ray that begins a particular type of event. Here, $N$ takes the value of either the incident hits, observed hits or observed true-energy hits to calculate the probability that a particular gamma-ray is in each of these scenarios.}
   \footnotesize
   \renewcommand\arraystretch{2}
   \begin{tabular}{llccc}
   \hline
   Scenario & Equation of Probability & Incident & Observed & True-energy\\
    &  & hits & hits & hits\\
   \hline
  Single hit & $Ne^{-2RT_{PP}^E}$ & 1 & 1 & 1\\
   \hline
   2 hits summed &  $Ne^{-2RT_{PP}^E}R T_{DT}^H$ & 2 & 1 & 0\\
   2 hits piled-up with no summing & $Ne^{-2RT_{PP}^E}(e^{-R T_{DT}^H}-e^{-R T_{PP}^E})$ & 2 & 2 & 2\\
   \hline
   3 hits all summed together & \(\displaystyle Ne^{-2 RT_{PP}^E}\frac{(R T_{DT}^H)^2}{2}\) & 3 & 1 & 0\\
   3 hits piled-up with hits 0 and 1 summed & $Ne^{-2RT_{PP}^E} R T_{DT}^H (e^{-R T_{DT}^H}-e^{-R T_{PP}^E})$ & 3 & 2 & 1\\
   3 hits piled-up with hits 1 and 2 summed & $Ne^{-2RT_{PP}^E} R T_{DT}^H (e^{-R T_{DT}^H}-e^{-R T_{PP}^E})$ & 3 & 2 & 1\\
   3 hits piled-up with no summing & $Ne^{-2RT_{PP}^E}(e^{-R T_{DT}^H}-e^{-R T_{PP}^E})^2$ & 3 & 3 & 3\\
   \hline
   4 hits all summed together & \(\displaystyle Ne^{-2 RT_{PP}^E}\frac{(R T_{DT}^H)^3}{6}\) & 4 & 1 & 0\\
   4 hits piled-up with hits 0 and 1, and &  &  & & \\
      hits 2 and 3 summed separately & \(\displaystyle Ne^{-2 RT_{PP}^E}(R T_{DT}^H)^2(e^{-R T_{DT}^H}-e^{-R T_{PP}^E})\) & 4 & 2 & 0\\
   4 hits piled-up with hits 0, 1 and 2 summed & \(\displaystyle Ne^{-2 RT_{PP}^E}\frac{(R T_{DT}^H)^2}{2}(e^{-R T_{DT}^H}-e^{-R T_{PP}^E})\) & 4 & 2 & 1\\
   4 hits piled-up with hits 1, 2 and 3 summed & \(\displaystyle Ne^{-2 RT_{PP}^E}\frac{(R T_{DT}^H)^2}{2}(e^{-R T_{DT}^H}-e^{-R T_{PP}^E})\) & 4 & 2 & 1\\
   4 hits piled-up with hits 0 and 1 summed & \(\displaystyle Ne^{-2 RT_{PP}^E}R T_{DT}^H(e^{-R T_{DT}^H}-e^{-R T_{PP}^E})^2\) & 4 & 3 & 2\\
   4 hits piled-up with hits 1 and 2 summed & \(\displaystyle Ne^{-2 RT_{PP}^E}R T_{DT}^H(e^{-R T_{DT}^H}-e^{-R T_{PP}^E})^2\) & 4 & 3 & 2\\
   4 hits piled-up with hits 2 and 3 summed & \(\displaystyle Ne^{-2 RT_{PP}^E}R T_{DT}^H(e^{-R T_{DT}^H}-e^{-R T_{PP}^E})^2\) & 4 & 3 & 2\\
   4 hits piled-up with no summing & $Ne^{-2RT_{PP}^E}(e^{-R T_{DT}^H}-e^{-R T_{PP}^E})^3$ & 4 & 4 & 4\\
   \hline
   \end{tabular}
\end{center}
\end{table*}

The probability for the various scenarios in which the true energy of the hit is detected, i.e. with no summing, is shown in Table \ref{tab:Probabilities_PU_summing}.

\begin{table*}[htp]
\begin{center}
\caption{\label{tab:Probabilities_PU_summing}Probabilities of true-energy hit detections at various incident gamma-ray rates. The last three columns capture all possibilities and therefore their sum is equal to 1.}
\begin{tabular}{ccccccccc}
\hline
Incident & $T^H_{DT}$ & $T^E_{dif}$ & \multicolumn{4}{c}{Number of pile-ups} & & Total full \\
rate (kHz) & $\mu$s & $\mu$s & 1 & 2 & 3 & $>=$4 & Summing & photopeaks \\
\hline
1 & 1.2 & 8 & 0.984 & 0.013 & 0.000 & 0.000 & 0.002 & 0.998 \\
5 & 1.2 & 8 & 0.923 & 0.061 & 0.003 & 0.000 & 0.012 & 0.988 \\
10 & 1.2 & 8 & 0.853 & 0.111 & 0.011 & 0.002 & 0.023 & 0.975 \\
20 & 1.2 & 8 & 0.731 & 0.180 & 0.034 & 0.012 & 0.044 & 0.944 \\
30 & 1.2 & 8 & 0.627 & 0.220 & 0.059 & 0.033 & 0.062 & 0.906 \\
40 & 1.2 & 8 & 0.539 & 0.239 & 0.082 & 0.065 & 0.075 & 0.860 \\
50 & 1.2 & 8 & 0.464 & 0.244 & 0.099 & 0.107 & 0.086 & 0.807 \\
\hline
50 & 1.2 & 4 & 0.681 & 0.165 & 0.030 & 0.021 & 0.104 & 0.876 \\
50 & 1.2 & 6 & 0.562 & 0.221 & 0.066 & 0.056 & 0.095 & 0.849 \\
50 & 1.2 & 8 & 0.464 & 0.244 & 0.099 & 0.107 & 0.086 & 0.807 \\
50 & 1.2 & 10 & 0.383 & 0.247 & 0.124 & 0.171 & 0.076 & 0.753 \\
\hline
50 & 1.0 & 8 & 0.462 & 0.252 & 0.106 & 0.107 & 0.072 & 0.821 \\
50 & 1.2 & 8 & 0.464 & 0.244 & 0.099 & 0.107 & 0.086 & 0.807 \\
50 & 1.4 & 8 & 0.466 & 0.236 & 0.093 & 0.107 & 0.099 & 0.794 \\
50 & 1.6 & 8 & 0.468 & 0.227 & 0.086 & 0.107 & 0.113 & 0.781 \\
50 & 1.8 & 8 & 0.470 & 0.219 & 0.080 & 0.106 & 0.126 & 0.768 \\
50 & 2.0 & 8 & 0.471 & 0.211 & 0.074 & 0.106 & 0.139 & 0.755 \\
\hline
\end{tabular}
\end{center}
\end{table*}

\bibliographystyle{elsarticle-num.bst}
\bibliography{GRIFFIN_DAQ_NIM}

\end{document}